\documentclass[12pt]{article} 
\usepackage{graphicx}
\usepackage{amsmath,amssymb}

\begin{document}

\baselineskip 21pt

\setcounter{totalnumber}{8}

\bigskip

\centerline{\Large \bf Young Star-Forming Complexes in the Ring}
\centerline{\Large \bf of the S0 galaxy NGC 4324}

\bigskip

\centerline{\large I.S. Proshina$^1$, A.V. Moiseev$^{2,1}$, and O.K. Sil'chenko$^1$}

\noindent

{\it Sternberg Astronomical Institute of the Lomonosov Moscow State University, Moscow, Russia}$^1$

\noindent

{\it Special Astrophysical Observatory of the Russian Academy of Sciences, Nizhnij Arkhyz, Russia}$^2$

\vspace{2mm}

\sloppypar 

\vspace{2mm}

\bigskip

{\small 

\noindent

We present the results of our study of starforming regions in the lenticular galaxy NGC 4324.
During a complex analysis of multiwavelength observational data -- the narrow-band emission-line images obtained with the 2.5-m telescope at the Caucasus Mountain Observatory of the Sternberg Astronomical Institute of the Moscow State University and the archival images in the broad bands of the SDSS, GALEX and WISE surveys -- we have detected young starforming complexes (clumps) located in the inner ring of the lenticular galaxy NGC 4324, and we have established a regular pattern 
of their distribution along the ring, which, nevertheless, changes with time (with age of starforming regions). We suggest several possible evolutionary paths of the lenticular galaxy NGC 4324, of which the accretion of gas-rich satellites or giant clouds (the so-called minor merging) is the most probable one.

}

\noindent


\clearpage

\section{INTRODUCTION}

Lenticular galaxies, by the definition of this morphological type, are generally believed to be disk galaxies without star formation. A deficit of gas in these early-type galaxies is again traditionally mentioned as being responsible for the absence of star formation in the disks of lenticular galaxies. However, deeper surveys in radio lines have recently shown that, in fact, quite often there is a cold gas in lenticular galaxies, both neutral hydrogen (Sage and Welch 2006; Serra et al. 2012) and molecular gas (Welch and Sage 2003; Welch et al. 2010), that can serve as a fuel for star formation. At the same time, the star formation being observed in some gas-rich lenticular galaxies is usually confined to ring structures (Pogge and Eskridge 1993; Salim et al. 2012) and, probably, may have a different trigger and slightly different physics than does the star formation in the arms of spiral galaxies. There is statistical evidence that the star formation in rings is much more efficient than the star formation in spiral arms (Kormendy and Kennicutt 2004).

\begin{figure*}[p]

\centering

\includegraphics[width=12cm]{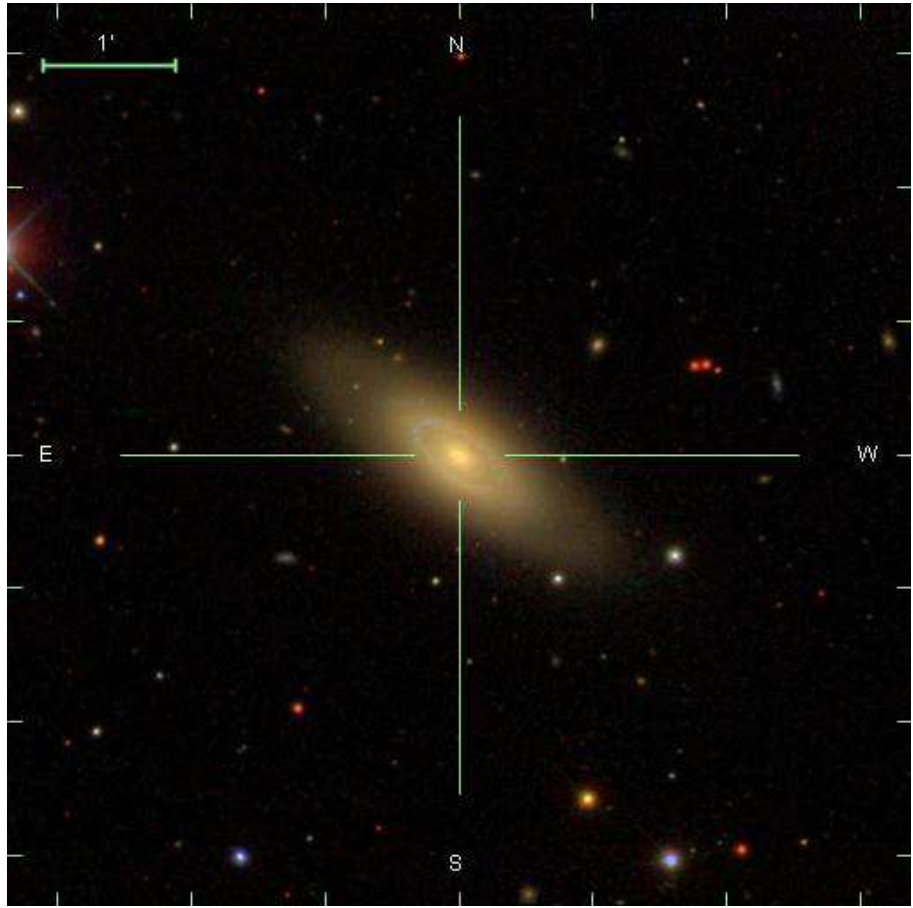}

\caption{The image of NGC 4324 in combined colors taken from the SDSS survey, DR9 (Ahn et al. 2012)}

\label{sdss}

\end{figure*}

The nearby early-type galaxy NGC~4324 being investigated here is remarkable for its bright blue ring (Fig. 1) embedded in a large-scale stellar disk typical for lenticular galaxies, having a reddish color and without distinct structural features, except the ring. The blue color of the ring points to current or recent star formation within it. The galaxy NGC~4324 was included into the sample of the ATLAS-3D project (Cappellari et al. 2011) and was investigated by means of panoramic spectroscopy. There are also photometric surveys that included NGC~4324. In the ARRAKIS atlas (Comer\'on et al. 2014), where the ring structures noticeable in the 3.6-$\mu$m and 4.5-$\mu$m bands are collected, it is classified as a galaxy with an inner ring, (L)SA(r)0$+$. A bar is suspected when inspecting the image left after the subtraction of the galaxy model constructed from the parameters derived during the decomposition of the image from the S4G survey (Sheth et al. 2010) given in ARRAKIS, although only the ring is pointed out among the distinguished features in the catalogue. However, the presence of a bar in this galaxy is also noted in the HyperLEDA database. According to this database, the galaxy under study is a member of the NGC~4303 group (Garcia 1993). Basic characteristics of the galaxy being investigated are given in Table~1.

\begin{table}

\caption[ ] {The main characteristics of NGC 4324}


\begin{flushleft}

\begin{tabular}{lc}

\hline\noalign{\smallskip}


Galaxy & NGC~4324 \\

Morphological type (NED$^1$) & SA(r)$0^+$ \\

$R_{25},\, ^{\prime \prime}$  (RC3$^2$) & 83 \\

$B_T ^0$ (LEDA$^3$) & 12.27$^m$ \\

$M_H$ (NED) & --23.43$^m$ \\

$(B-V)_e$ (LEDA) & 0.92$^m$ \\

PA$_{phot}$ (NED) & $52^{\circ}$ \\

Inclination i$_{phot}$ (NED) & $63^{\circ}$ \\ 

$V_r$, km/s (NED) & $1667\pm 3$ \\

Distance$^4$, Mpc & 26.2 \\

$M_B$ & --19.82$^m$ \\

Metric scale & 127 pc$/^{\prime \prime}$  \\

\hline

\multicolumn{2}{l}{$^1$\rule{0pt}{11pt}\footnotesize

NASA/IPAC Extragalactic Database}\\

\multicolumn{2}{l}{$^2$\rule{0pt}{11pt}\footnotesize

Third Reference Catalogue of Bright Galaxies, de Vaucouleurs et al. (1991)} \\

\multicolumn{2}{l}{$^3$\rule{0pt}{11pt}\footnotesize

Lyon-Meudon Extragalactic Database}\\

\multicolumn{2}{l}{$^4$\rule{0pt}{11pt}\footnotesize

Cosmicflow-2, Tully et al.(2013)} \\

\end{tabular}

\end{flushleft}

\end{table}

There is a lot of gas present in this early-type galaxy. For example, an estimate of the molecular hydrogen mass is given in the paper of the ATLAS-3D project by Young et al. (2011): $\log M(H_2) = 7.69 \pm 0.05$. The maps of the CO distribution and kinematics are shown in the detailed study of the molecular gas within the same project (Alatalo et al. 2013): molecular hydrogen is concentrated in a ring closely following the optically visible ring in morphology and kinematics, and its distribution coincides with the stellar ring and the ionized gas ring. The molecular hydrogen mass in this paper is estimated to be $\log M(H_2) = 7.97 \pm 0.02$. The coincident kinematic position angles of the lines of nodes of the stellar disk, $\phi _{star} = 238^{\circ} \pm 1^{\circ}$, the molecular, $\phi _{mol} = 232.0^{\circ} \pm 1.8^{\circ}$, and ionized, $\phi _{ion} = 239.0^{\circ} \pm 6.8^{\circ}$, gas obtained in ATLAS-3D (Krajnovi\'c et al. 2011; Davis et al. 2011a) suggest that the morphology and rotation of the molecular and ionized gas and the stellar disk (in the vicinity of the ring) coincide. The normalized surface brightness profile of the molecular gas emission with its peak at a radius of $20^{\prime \prime}$, almost at the radius of the stellar ring, given the poorer spatial resolution of the CO observations, is shown in Davis et al. (2011b). 

The earliest neutral hydrogen mass estimates are given in Krumm and Salpeter (1979) and Giovanardi et al. (1983); they are $6 \times 10^8$~M$_{\odot}$ and $5.1 \times 10^8$~M$_{\odot}$, respectively. According to Cortese and Hughes (2009), the galaxy under study contains $6.76 \times 10^8$~M$_{\odot}$ of neutral hydrogen. As regards the distribution of neutral hydrogen in the galaxy, it is reported in Duprie and Schneider (1996) that neutral hydrogen is detected at distances up to two optical radii, i.e., it is distributed over the whole disk and beyond rather than concentrated only in the ring, consistent with the position-velocity map in Hoffman et al. (1989).

It is not surprising that with such amount of gas, with 50--100 million solar masses of H$_2$ being confined to a narrow range in radius, young stars are formed in the ring of the galaxy NGC~4324. We have already studied NGC 4324 spectroscopically: these were observations with a long slit at the South African Large Telescope (SALT) (Proshina et al. 2019) and observations with a scanning Fabry-Perot interferometer at the 6-m BTA telescope (Sil'chenko et al. 2019). Our previous study (Proshina et al. 2019) showed the presence of bright emission lines in the spectrum of this galaxy and nonuniformity of the distribution of starforming sites along the slit; the H$\alpha$ intensity peaks at the ring at a distance of $23^{\prime \prime}$ from the center. The ring of the galaxy is also excellently seen in the ultraviolet in the data by the GALEX space telescope survey (Bouquin et al. 2018). We decided to investigate the pattern of star formation in the ring of NGC~4324 by combining the GALEX data and our own narrow-band photometric imaging of the galaxy in the ionized gas H$\alpha$ emission that characterizes the current star formation rates (SFRs) on a timescale up to 10~Myr. In the next section we describe our observations, then list our results on the characteristics of starforming regions in the ring, and then present a discussion of our results and conclusions.

\section{OBSERVATIONS AND DATA ANALYSIS}

We decided to take full images of the galaxy NGC~4324 in narrow photometric bands centered on the H$\alpha$ and [NII]$\lambda$6583 emission lines with the MaNGaL instrument -- a mapper with a tunable filter. A detailed description of the instrument is given in Moiseev et al. (2020). The small bandwidth of the tunable filter, 13~\AA, allows an image in each emission line to be taken separately. This, in turn, allows us to compare the fluxes in emission lines and to determine the gas excitation mechanisms based on diagnostic diagrams. The observations were carried out on April 17, 2018, with the 2.5-m telescope at the Caucasus Mountain Observatory (Shatsky et al. 2020). The total exposure time was 1500~s for the H$\alpha$ line image, 3000~s for the [NII]$\lambda$6583 line image, and 1500~s for the image in the red continuum adjacent to the lines. The scale of all images was 0.66 arcsec per pixel ($2 \times 2$ binning); the seeing allowed maps with a spatial resolution of 1.5 arcsec to be constructed. We described the technique for subtracting the continuum images from the narrow-band MaNGaL emission-line data to obtain ''pure'' emission-line images, including the calibration of the observed fluxes in energy units and the correction of the images for the overlapping wings of close emission lines due to the finite spectral resolution of the instrument, in Sil'chenko et al. (2020).

In addition, for a more comprehensive analysis of the star formation process we used the ultraviolet (UV) images of the galaxy under study obtained with the GALEX space telescope retrieved from the public Mikulski Archive for Space Telescopes (MAST), and its optical images from SDSS DR9 (Ahn et al. 2012), as well as its image in the W4 band at 22 $\mu$m obtained with the WISE space telescope from the public NASA/IPAC archive. Table~2 gives the identifiers of the observing programs, the dates of observations, and the exposure times of the UV images of NGC~4324 used by us.

\begin{table}

\caption[ ] {Listing of the GALEX observations of NGC 4324}


\begin{flushleft}

\begin{tabular}{llcc}

\hline\noalign{\smallskip}


Band & ProgID & The date of the observations & Exposure time, s \\

FUV &    AIS\,228\,0001\,sg28 &   March 31, 2004 &  106.5  \\

NUV &    AIS\,228\,0001\,sg27  &  March 31, 2004  &  200  \\

NUV &    MISGCSN1\,13360\,0229 &  April 1, 2011 & 1739.7  \\

NUV &    GI6\,001033\,GUVICS033 & March 18, 2010  &  1668.2  \\

\hline

\end{tabular}

\end{flushleft}

\end{table}

\begin{figure*}[p]

\centering

   \vspace{0.5cm}

   \centerline{

   \includegraphics[width=0.45\textwidth]{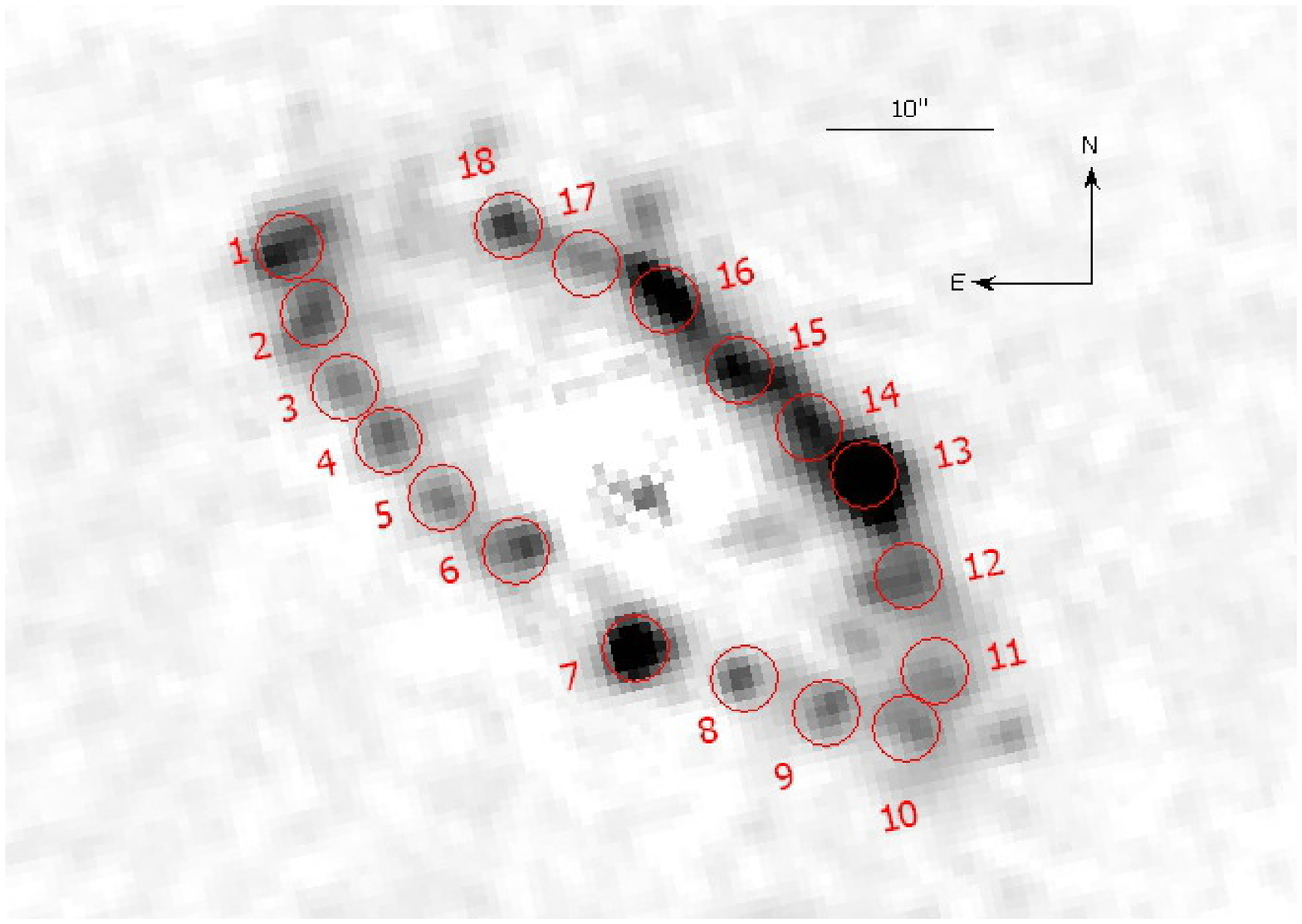}

   \includegraphics[width=0.45\textwidth]{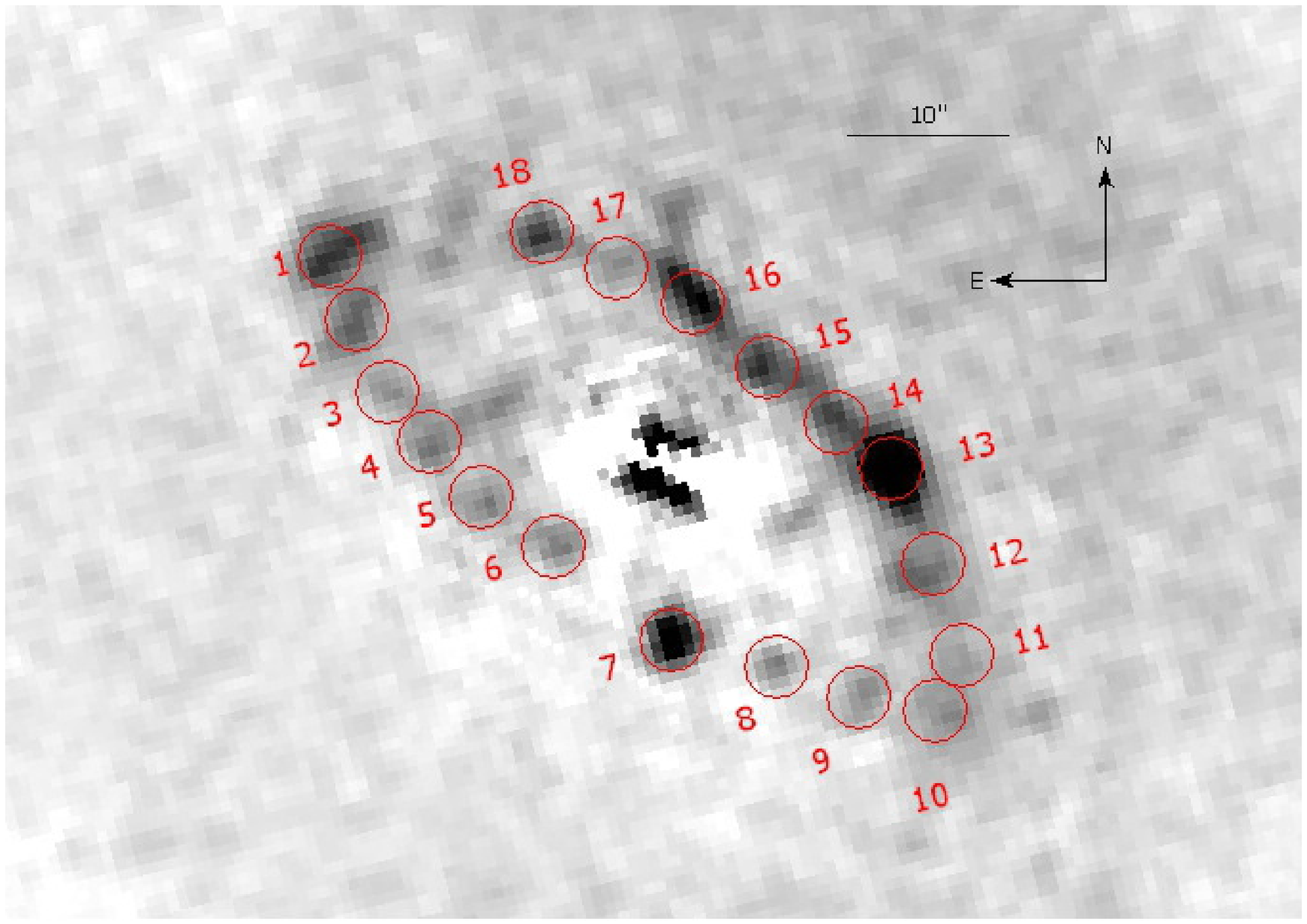}

   }

   \centerline{

   \includegraphics[width=0.45\textwidth]{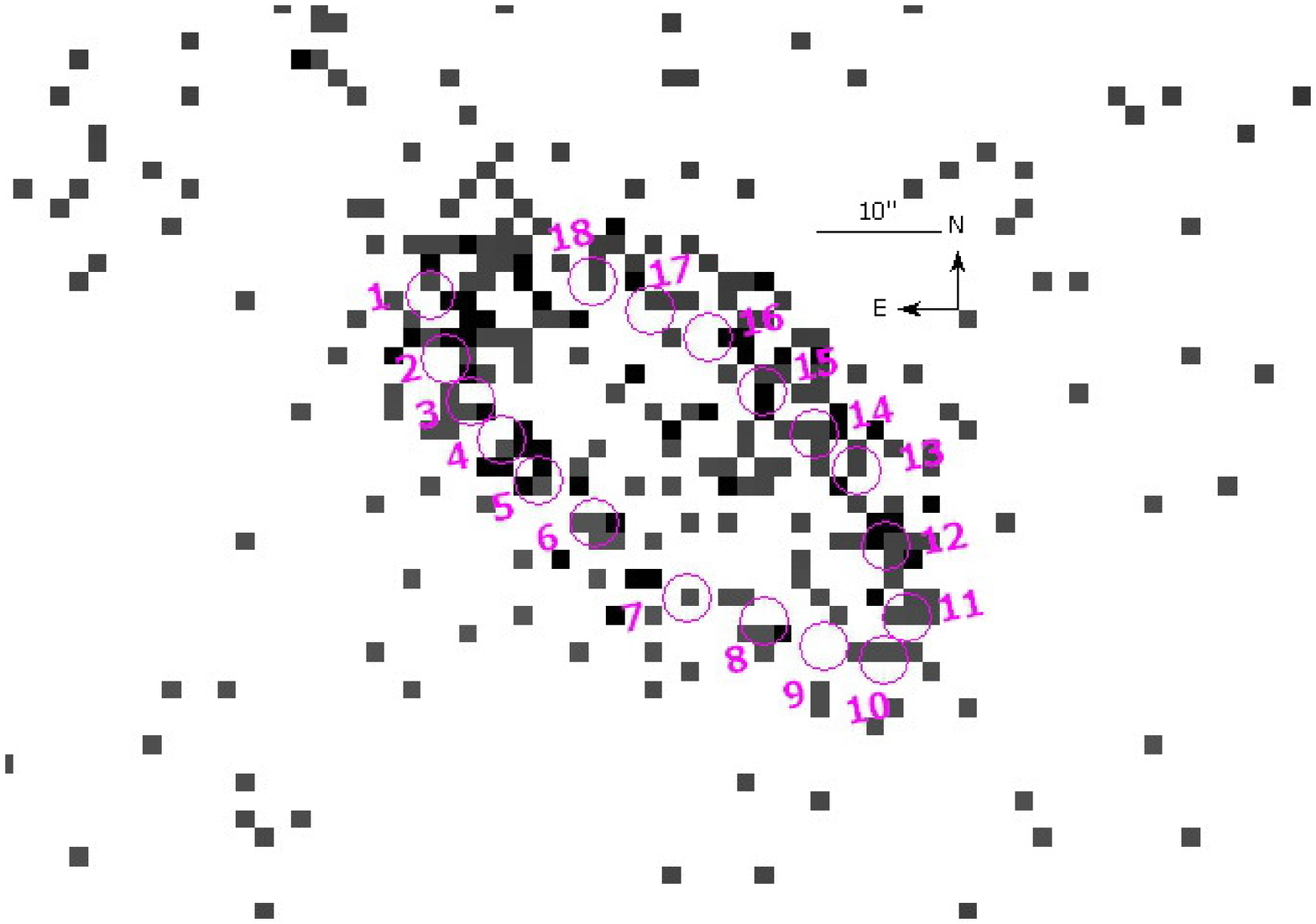}

   \includegraphics[width=0.45\textwidth]{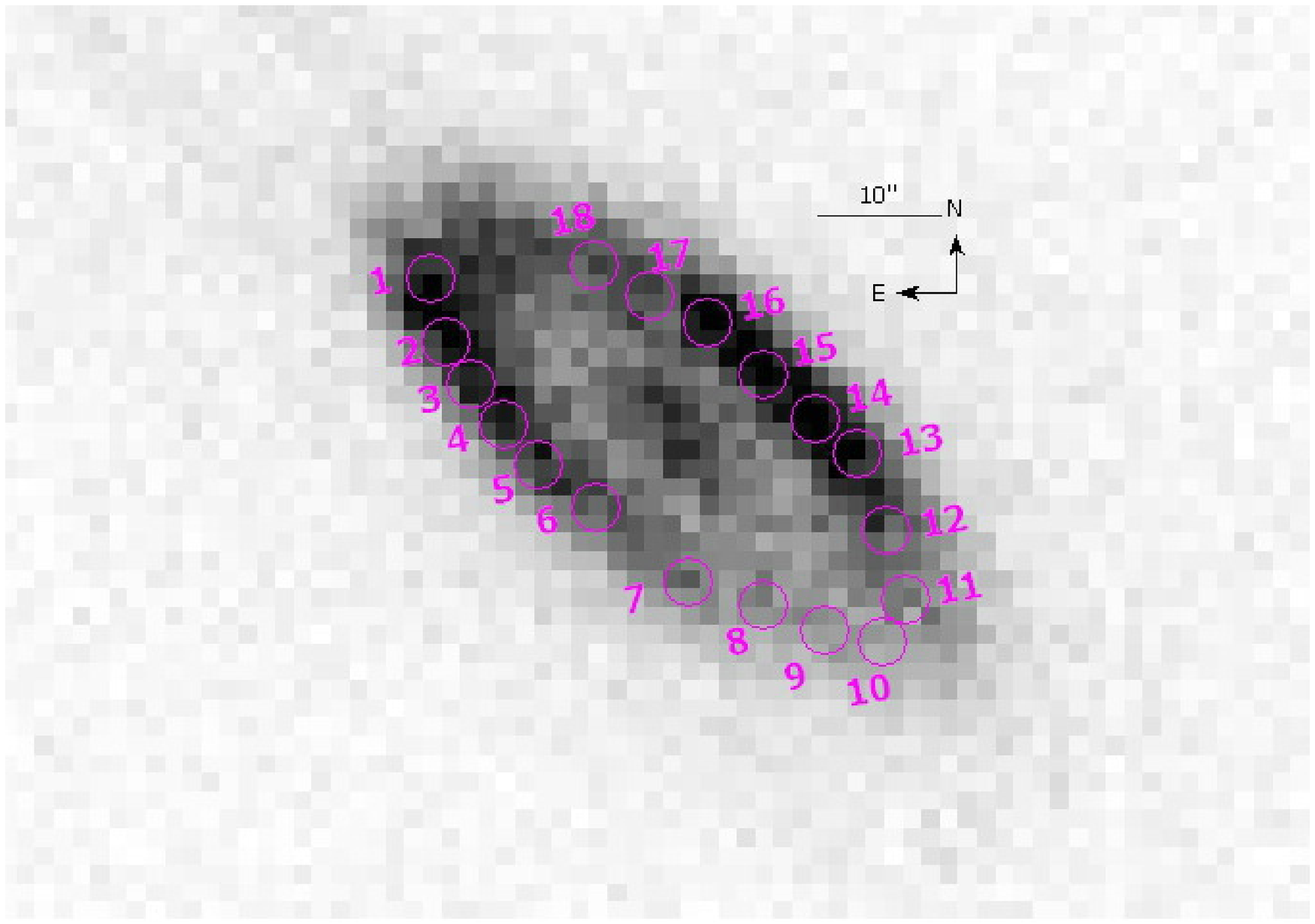} }

   \caption{The ring of NGC 4324: the upper raw presents the data by MaNGaL, in the narrow bands centered onto emission lines H$\alpha$ (left) and [NII]$\lambda$6583 (right), the bottom raw -- the data by GALEX, in the bands FUV (1500\AA) and NUV (2300\AA).}

\label{ring}

\end{figure*}

Compact emission-line regions in the ring (clumps) are clearly seen on the H$\alpha$ and [NII]$\lambda$6583 intensity maps with the subtracted continuum. The typical size of the clumps turned out to be about $4^{\prime \prime}$, or 0.5 kpc. The emission-line flux was integrated in apertures of such a size. Figure~2 (upper row) shows the galaxy's images in the narrow H$\alpha$ and [NII]$\lambda$6583 emission lines with the apertures centered on the clumps and numbered along the ring. The same apertures were also superimposed on the UV images in the FUV and NUV bands and on the WISE/W4 image. The fluxes of the clumps were measured in the specified apertures. For the UV images the fluxes were converted to magnitudes using the calibration equations from Morrissey et al. (2007). The derived magnitudes were corrected for dust absorption in our Galaxy using the photometric calibrations from the NED ($A_B=0.087$ for NGC~4324). The UV fluxes were corrected for intrinsic absorption in the galaxy using the infrared dust radiation estimates based on the WISE/W4 data.

\section{THE RESULTS}

\subsection{The gas excitation diagnostics}

Table~3 gives the measured emission-line fluxes and metallicity estimates for the ionized gas in the regions where we detect the dominant contribution of star formation to the gas excitation. The emission-line ratio [NII]$\lambda$6583/H$\alpha$ was corrected for the overlapping of the line wings in the MaNGaL band, according to the calibration by Moiseev et al. (2020).

\begin{table*}

\caption[ ] {The characteristics of the clumps in the ring of NGC 4324}


\begin{flushleft}

\begin{tabular}{rcccccc}

\hline\noalign{\smallskip}


No.  & Flux in H$\alpha$, &  $\Sigma$(H$\alpha$), & EW (H$\alpha$), &

$\log \frac {\mbox{[NII]}\lambda 6583}{\mbox{H}\alpha}$ & \multicolumn{2}{c}{$12+\log \mbox{(O/H)} $} \\

  & erg/s/cm$^2$ & erg/s/kpc$^2$  & \AA\  &   & Pettini, Pagel(2004)  & Marino$+$(2013) \\

\hline\noalign{\smallskip}

1 & 3.60E-15 & 1.69E+39 & $7.87\pm 0.20$ & $-0.30\pm 0.01$ &  &  \\

2 & 2.74E-15 & 1.32E+39 & $5.23\pm 0.20$ & $-0.32\pm 0.01$ &  &  \\

3 & 1.94E-15 & 1.05E+39 & $2.62\pm 0.20$ & $-0.35\pm 0.01$ &      &      \\

4 & 2.17E-15 & 1.16E+39 & $2.08\pm 0.14$ & $-0.31\pm 0.01$ &      &      \\

5 & 1.99E-15 & 1.09E+39 & $1.62\pm 0.10$ & $-0.31\pm 0.01$ &      &      \\

6 & 2.66E-15 & 1.36E+39 & $1.85\pm 0.11$ & $-0.54\pm 0.01$ &     &      \\

7 & 5.34E-15 & 2.41E+39 & $5.13\pm 0.10$ & $-0.49\pm 0.01$ & 8.62 & 8.51 \\

8 & 1.87E-15 & 1.00E+39 & $2.60\pm 0.10$ & $-0.49\pm 0.01$ &      &     \\

9 & 2.20E-15 & 1.10E+39 & $5.63\pm 0.15$ & $-0.42\pm 0.01$ & 8.66 & 8.55 \\

10 & 2.05E-15 & 0.96E+39 & $8.82\pm 0.20$ & $-0.32\pm 0.01$ &  &  \\

11 & 2.18E-15 & 1.04E+39 & $8.63\pm 0.20$ & $-0.41\pm 0.01$ & 8.67 & 8.55 \\

12 & 2.99E-15 & 1.43E+39 & $7.69\pm 0.20$ & $-0.40\pm 0.01$ &  &   \\

13 & 1.44E-14 & 5.97E+39 & $30.6\pm 0.2$ & $-0.54\pm 0.005$ & 8.59 & 8.49 \\

14 & 4.07E-15 & 1.91E+39 & $6.59\pm 0.16$ & $-0.41\pm 0.01$ & 8.67 & 8.55 \\

15 & 3.94E-15 & 1.91E+39 & $4.58\pm 0.10$ & $-0.42\pm 0.01$ & 8.66 & 8.55 \\

16 & 4.66E-15 & 2.14E+39 & $5.93\pm 0.20$ & $-0.43\pm 0.01$ & 8.65 & 8.54 \\

17 & 1.89E-15 & 1.05E+39 & $2.57\pm 0.10$ & $-0.32\pm 0.01$ &      &     \\

18 & 2.58E-15 & 1.30E+39 & $3.69\pm 0.15$ & $-0.22\pm 0.01$ &   &  \\

\hline

\end{tabular}

\end{flushleft}

\end{table*}

There are several methods by which the gas excitation mechanism can be determined. The
first criterion involves the diagnostic diagrams proposed by Baldwin et al. (1981), the
so-called BPT diagrams, which allow the mechanisms of gas excitation by shock waves or 
active nuclei to be separated from the excitation by UV radiation from young OB stars 
based on emission-line intensity ratios. By using the model calculations of these 
diagrams from Kewley et al. (2006), we adopted the threshold value of $\log (\mbox{[NII]}\lambda 6583 / \mbox{H}\alpha )=-0.41$ below which the gas may be 
considered as being excited exclusively by radiation from young stars. At the same 
time, we adopt $\log (\mbox{[OIII]}\lambda 5007 / \mbox{H}\beta ) = 0.10$ in 
the ring, according to the measurement of the longslit spectrum of NGC~4324 in our 
previous paper (Proshina et al. 2019). Another criterion proposed by Zhang et al.
(2017) is related to the H$\alpha$ surface brightness at which the emission is assumed 
to be produced by the gas ionization by radiation from young stars: $\Sigma (\mbox{H}\alpha) > 10^{39}$ erg s$^{-1}$ kpc$^{-2}$. 
Yet another criterion for the identification of starforming regions is related to the 
equivalent width of the H$\alpha$ emission line. For example, Binette et al. (1994) and 
Cid Fernandes et al. (2011) suggested a threshold value of EW(H$\alpha$) = 3~\AA ; 
values below 3~\AA\ may be produced by diffuse ionized gas (DIG) regions. However, 
Lacerda et al. (2018) suggested a different threshold value of EW(H$\alpha$), 14~\AA. 
The ambiguity of this criterion stems from the fact that in the case of projecting the 
starforming regions onto a bright underlying structure (for example, when the 
starforming regions are not far from a galactic bulge), the contribution of this 
underlying structure should be taken into account, that reduces the reliability of 
assigning the measured emission-line equivalent widths precisely to the distinguished 
gas emission region. In the case of our galaxy, the ring of NGC~4324 is the inner one, 
with a radius $\sim 3$~kpc, and, therefore, the contribution of the underlying stellar 
population of the galaxy should be taken into account, which we did. For confident 
identification of the clumps with regions of current star formation we apply all three criteria.

It is for the clumps, where the gas is excited by young stars, that we can estimate the 
gas metallicity from the ratio of the fluxes in the [NII]$\lambda$6583 and H$\alpha$ 
emission lines using, for comparison, two calibrations from Pettini and Pagel (2004) 
and Marino et al. (2013). Figure~3 shows how the gas metallicity changes from the clump 
to the clump with their SFR. When considering the ''instantaneous'' SFR determined
from the H$\alpha$ flux, this dependence for SFRs above $10^{-2}\,M_{\odot}$~yr$^{-1}$ kpc$^{-2}$ is the inverse one.
When averaging the SFRs on a timescale of hundreds Myr, using the NUV-flux indicator, 
this anticorrelation is washed out. This can be probably explained by a short duration 
of local starbursts: as the chemical evolution in the clump advances, the gas is 
locally depleted, while the gas metallicity reaches saturation (Ascasibar et al. 2015). 
The solar metallicity is the final point of the chemical evolution: the gas
metallicity reaches a plateau near the solar value, when the density of the young 
stellar population begins to exceed locally the gas density (Ascasibar et al. 2015). 
Note that the clumps with a low SFR in Fig.~3 do not fall on the extension of the above 
dependence. Most likely, the reason is that, in fact, the emission of these clumps is 
largely produced by shock processes, and these clumps were attributed to the
starforming ones formally, because their H$\alpha$ surface brightness exceeds only 
marginally the critical value of $10^{39}$ erg s$^{-1}$ kpc$^{-2}$ suggested by Zhang et al. (2017).

\begin{figure*}[p]

\centering

   \vspace{0.5cm}

   \centerline{

   \includegraphics[width=0.45\textwidth]{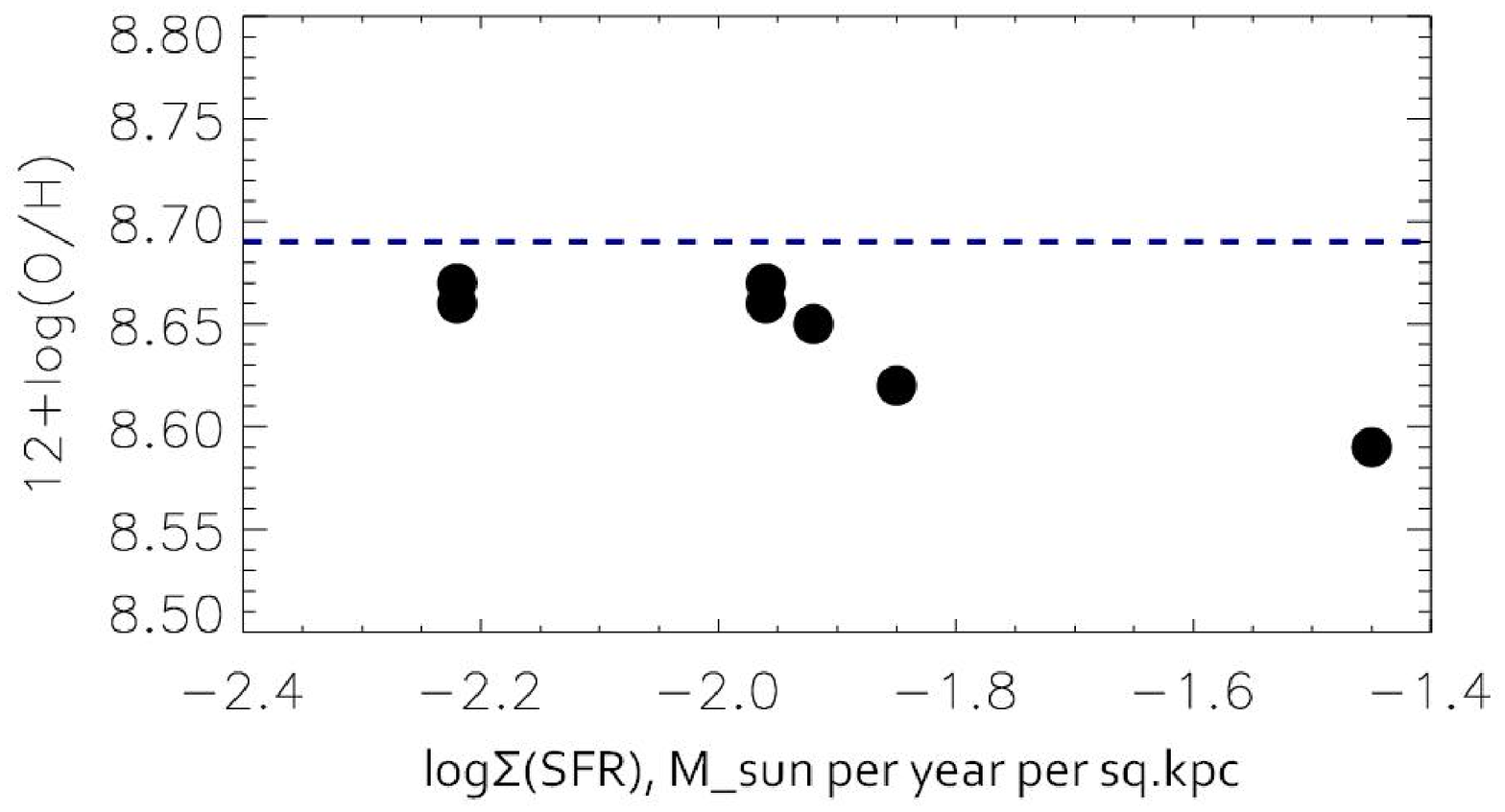}

   \includegraphics[width=0.45\textwidth]{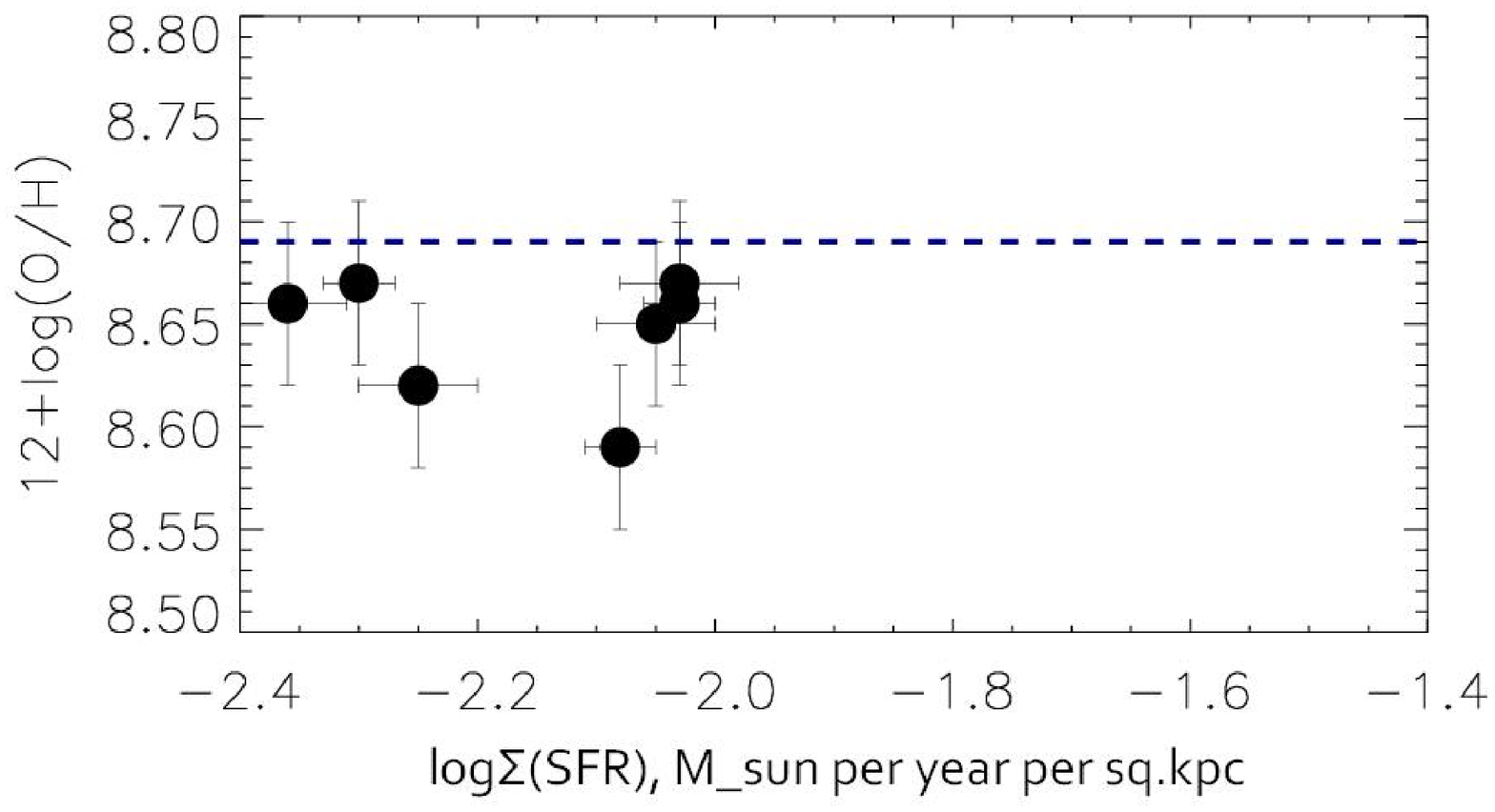}

   }

\caption{The (anti-)correlation between the gas metallicity of the clumps and the local star formation rate: the left -- the correlation between the oxygen abundance in the ionized gas according the Pettini and Pagel (2004) calibration and the instant star formation rate; the right -- the metallicity against the star formation rate averaged over the timescale of 200 Myr (derived from the GALEX/NUV flux). The horizontal dashed line marks the solar metallicity.}

\label{sfrmet}

\end{figure*}

\subsection{Star formation rate estimates}

The SFRs used to construct the dependence in Fig.~3 were calculated from the fluxes 
that we have measured based on the MaNGaL map in the H$\alpha$ emission line; the 
calibration of SFR through H$\alpha$ is from the review by Kennicutt and Evans (2012). 
The H$\alpha$ fluxes were corrected for the intrinsic dust absorption in the galaxy; to 
make the correction for dust, we used the galaxy's image taken with the WISE space 
telescope in the W4-band (22 $\mu$m) and retrieved by us in the open NASA/IPAC archive. 
The H$\alpha$ emission is a star formation tracer on a short timescale which does not 
exceed 10~Myr. The review by Kennicutt and Evans (2012) also provides the calibrations 
to calculate the SFRs from the measured UV fluxes, which are SFR tracers on longer 
timescales -- from 100 to 200~Myr. We carried out these calculations using the FUV and
NUV images of NGC~4324 from the GALEX space telescope (also by taking into account the 
UV dust absorption in the galaxy under study using the WISE image in the W4-band). The 
SFR estimates obtained through the NUV fluxes from three observing programs (AIS, MIS, 
and GI) are quite consistent; the availability of three independent measurements in the 
NUV band allows us both to calculate the mean values and to estimate the errors of the 
derived SFRs. The plot of the SFR density variation along the ring of NGC~4324 (Fig.~4) 
turns out to be very curious: a drop in the SFR surface density derived from the FUV 
flux is observed just where intense star formation from the flux in the H$\alpha$ 
emission line is now observed (clumps 7 and 13). According to the remark by Calzetti 
(2013), this is typical for a very short timescale, when the star formation lasts at 
present no more than 2~Myr. In this case, the estimates obtained from the UV fluxes 
based on the calibrations by Kennicutt and Evans (2012) should be multiplied by a 
factor of 3.45. Thus, we conclude that the starbursts in these clumps have begun quite 
recently (within 10~Myr). The stars that could be formed during a previous starburst ($\sim$200~Myr ago) cannot contribute now to the FUV luminosity having already exploded, 
or their luminosity peak has now shifted to the NUV, leading to the observed dip in the 
FUV (clumps 7, 13, 16, 17, 18, 1). It is worth noting that the curve tracing the SFRs 
calculated from the NUV flux is almost horizontal and, therefore, this indicator, the 
NUV data, is better to determine the mean SFR along the ring. It can be assumed that 
the star formation began first in clumps 2-3-4, then in 5-6, 14-15, and 8-12, 
subsequently in 16-17-18-1, then in the clump 7 and in the clump 13 -- the most recent 
starburst (possibly a recurrent one, 200~Myr after the previous one).

\begin{figure*}[p]

\centering

\includegraphics[width=12cm]{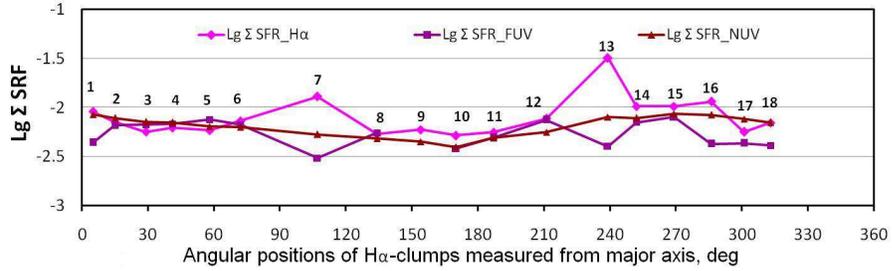}

\caption{The star formation rate density variations along the ring}

\label{sfralong}

\end{figure*}

\subsection{Orientation of the Gaseous Disk of NGC 4324 from the Ionized Gas Rotation Velocity Field}

\begin{figure*}

\centerline{

\includegraphics[height=5cm]{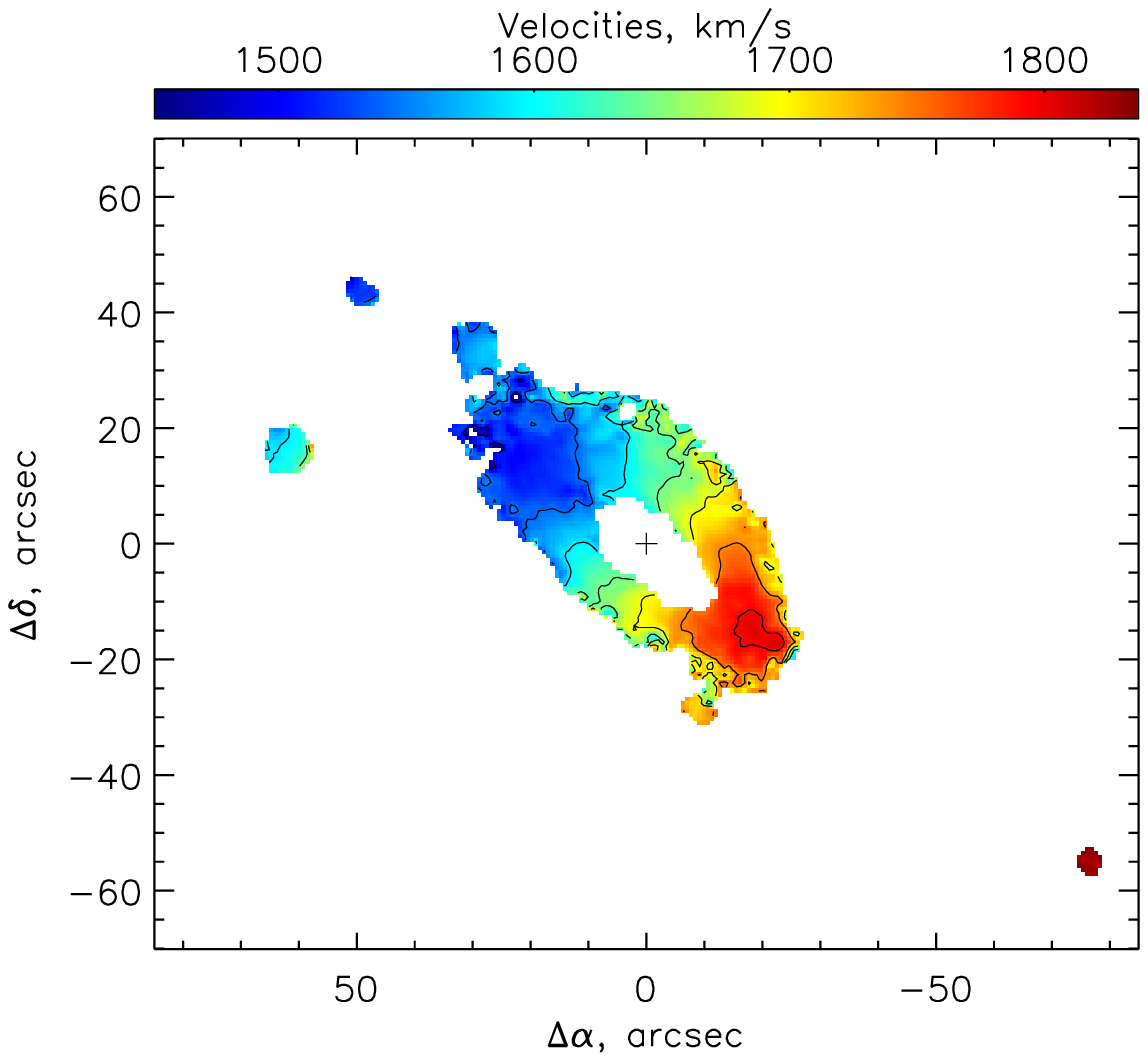}

\includegraphics[height=5cm]{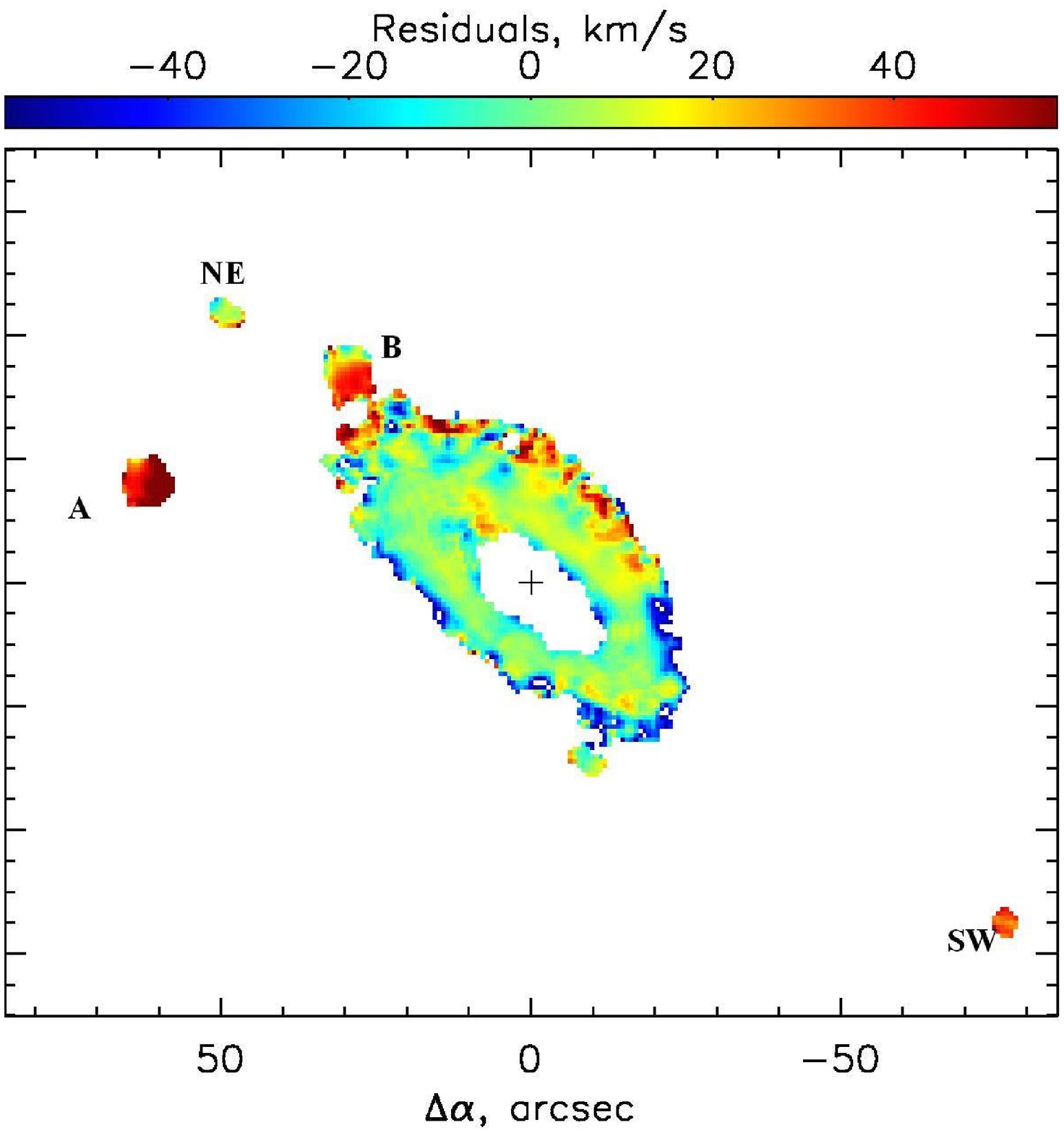}

\includegraphics[height=5cm]{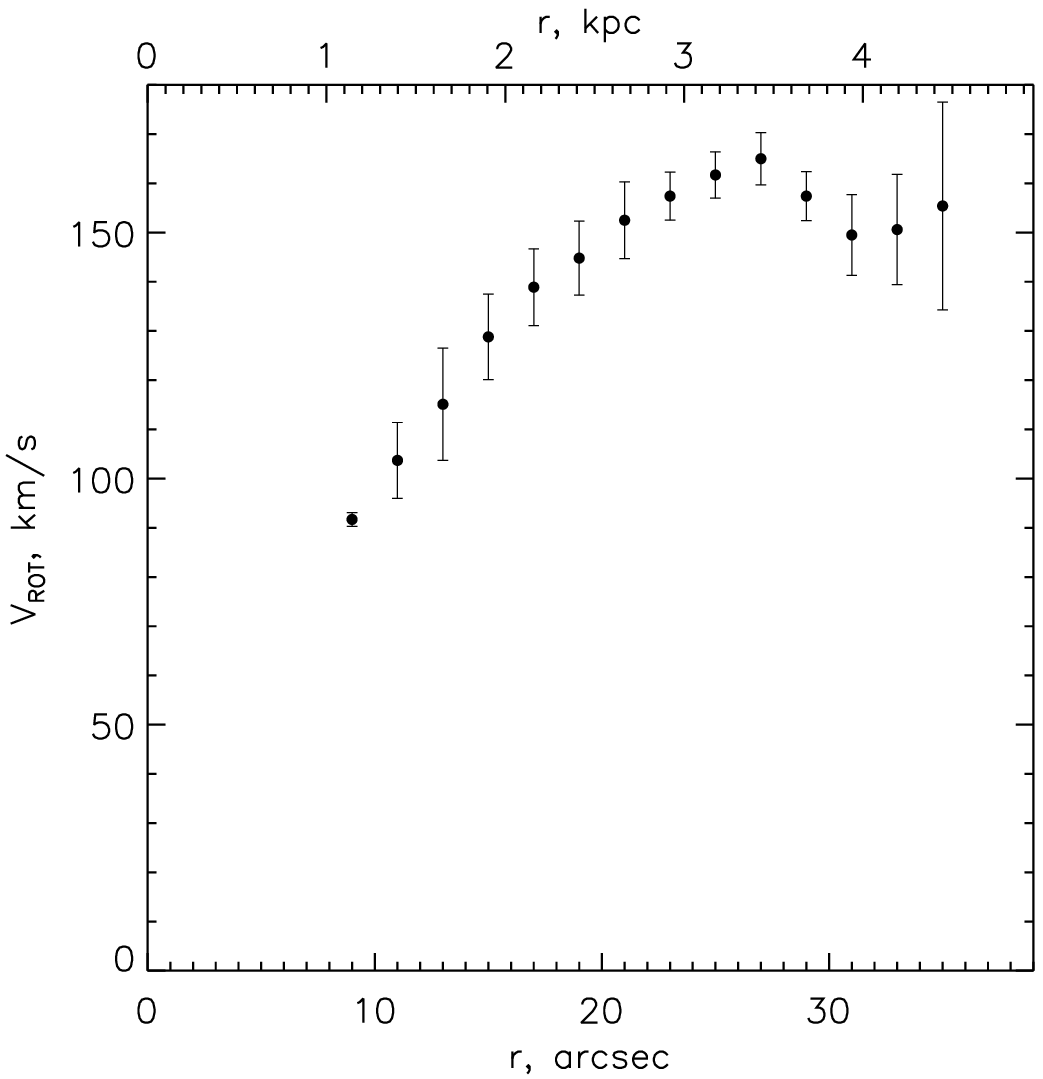}

}

\caption{The kinematics of the ionized gas from the Fabry-Perot data in the emission line H$\alpha$. From left to right: the line-of-sight velocity field, the residual velocities after subtracting the model of circular rotation, the adopted rotation curve.}

\label{ifpdata}

\end{figure*}

In our previous paper (Sil'chenko et al. 2019) we presented, among the data for a 
sample of 18 lenticular galaxies, our panoramic spectroscopy for NGC~4324 in the H$\alpha$ emission line obtained with a scanning Fabry-Perot interferometer of the 6-m 
telescope of the Special Astrophysical Observatory. A more detailed analysis of the 
data presented now in Fig.~5 has shown that the gas is mostly involved in circular 
rotation being confined to a plane inclined to our line of sight at an angle of
$65^{\circ} \pm 3^{\circ}$ and with a line of nodes oriented in the sky at a position 
angle of $PA = 235^{\circ} \pm 3^{\circ}$ (the radial range for our analysis is $10^{\prime \prime} - 35^{\prime \prime}$). This inclination, in principle, is 
consistent with the inclination estimated from our isophotal analysis for the inner 
galactic stellar disk (Proshina et al. 2019), $63^{\circ}$, as it must be in the case
of circular gas rotation in the main stellar disk plane. However, emission-line clumps 
whose velocities differ significantly, up to 70 km s$^{-1}$, from the model of 
extrapolated circular rotation with a flat curve V(R), are visible in the outer disk 
regions and in the immediate vicinity of the galaxy. Below in the Discussion we 
will use these data to justify the hypothesis of external gas accretion onto NGC~4324.

\subsection{Regularity of the Distribution of Clumps along the Ring}

\begin{figure*}[p]

\centering

   \vspace{0.5cm}

   \centerline{

   \includegraphics[width=0.4\textwidth]{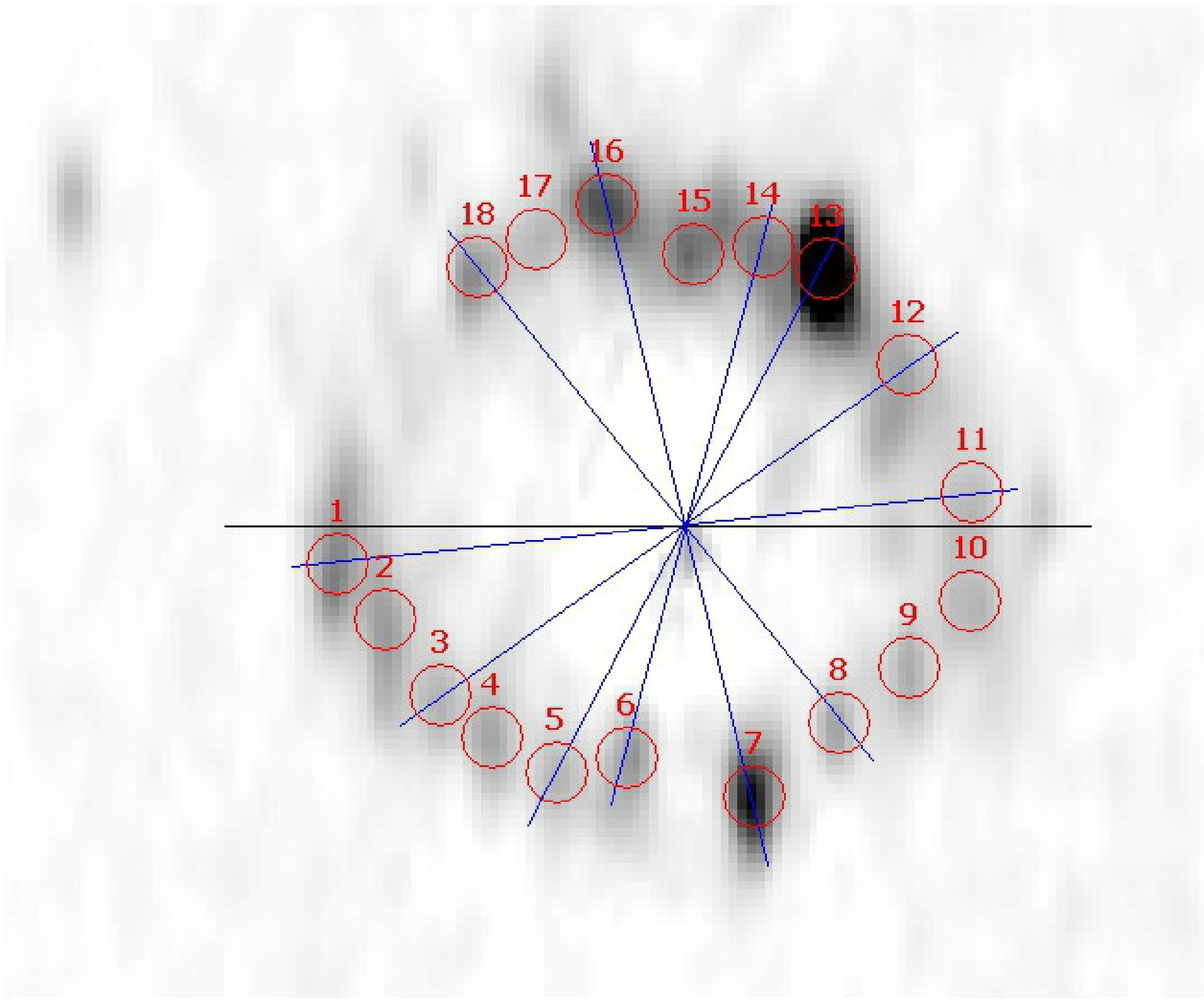}

   \includegraphics[width=0.4\textwidth]{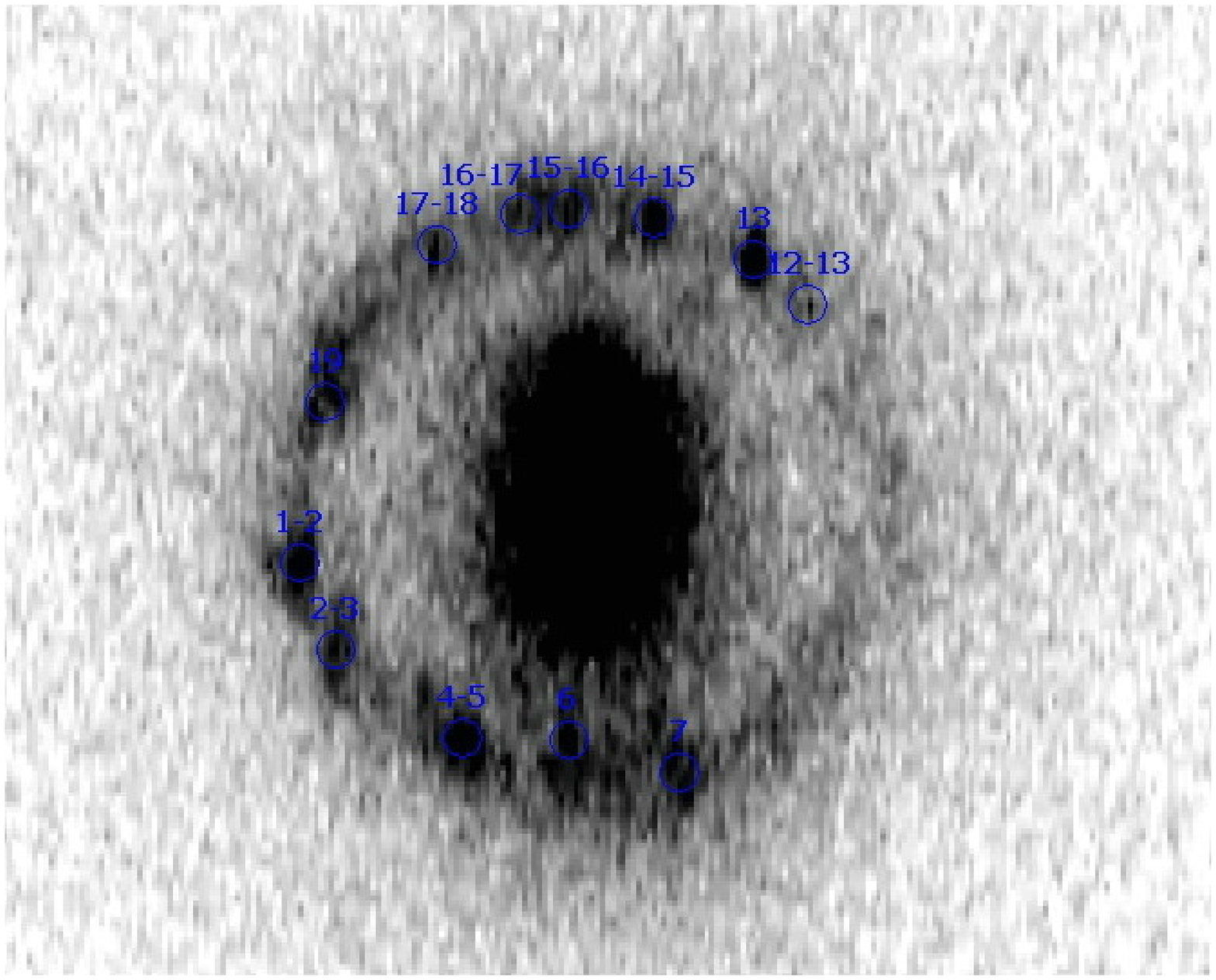}

   }

   \centerline{

   \includegraphics[width=12cm]{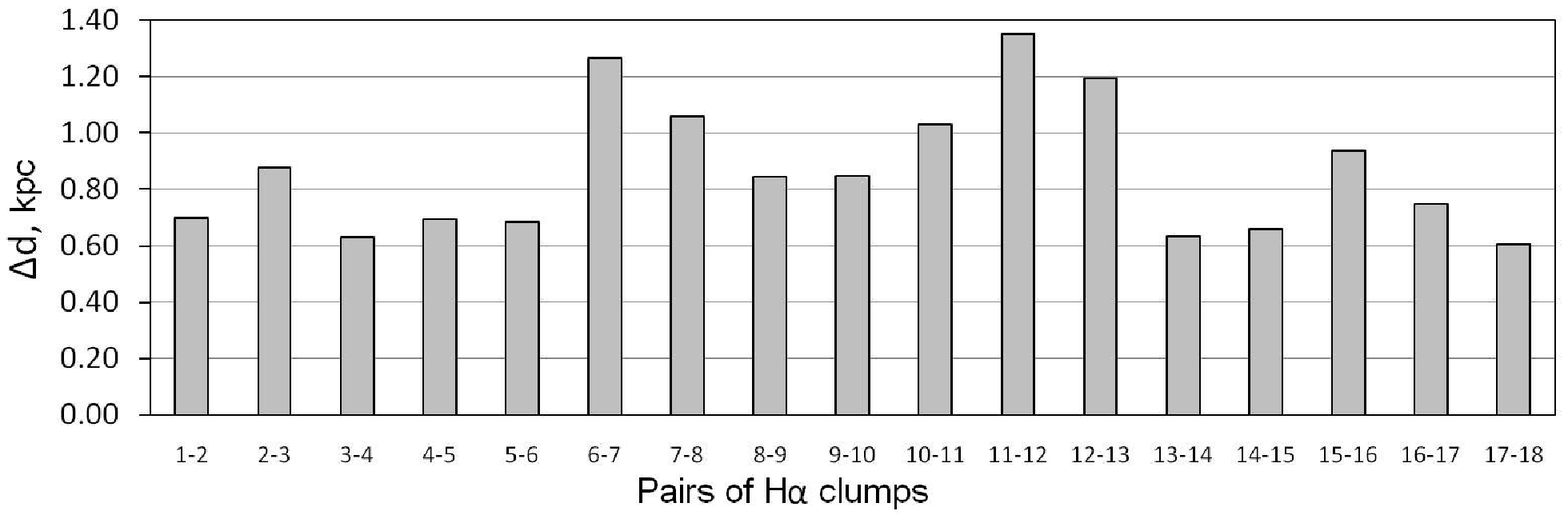}}

   \centerline{

   \includegraphics[width=12cm]{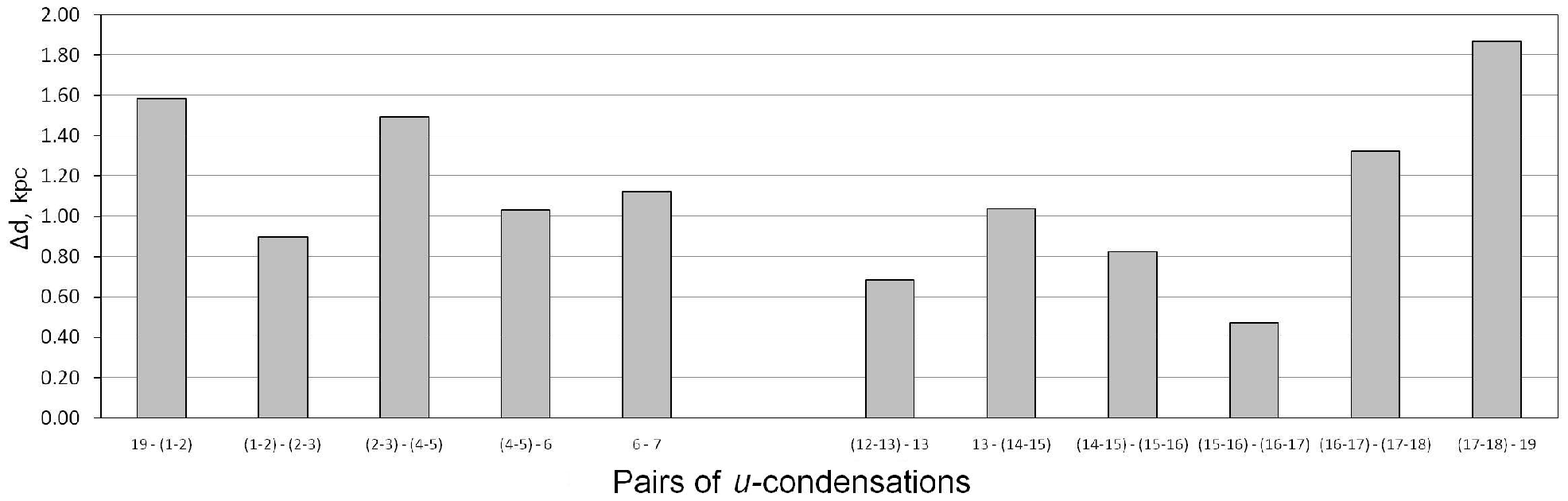} }

   \caption{The ring structure in the emission line H$\alpha$ (left) and in the $u$-band (right): the upper raw -- the deprojected ring images, at the bottom -- the separations between the adjacent clumps in the emission line H$\alpha$ and in the $u$-band.}

\label{ringdepro}

\end{figure*}

\begin{figure*}[p]

\centering

   \vspace{0.5cm}

   \centerline{

   \includegraphics[width=0.45\textwidth]{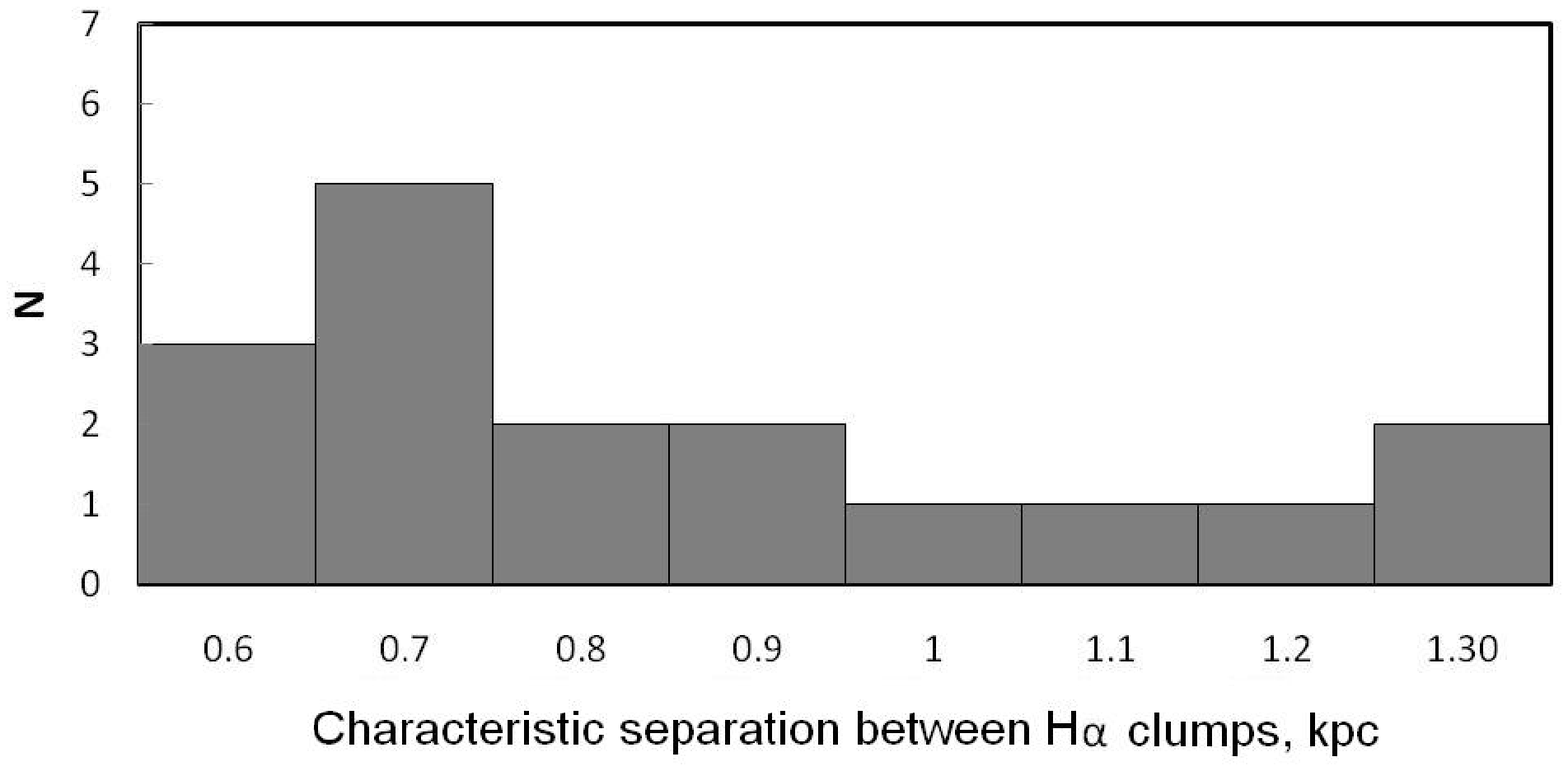}

   \includegraphics[width=0.45\textwidth]{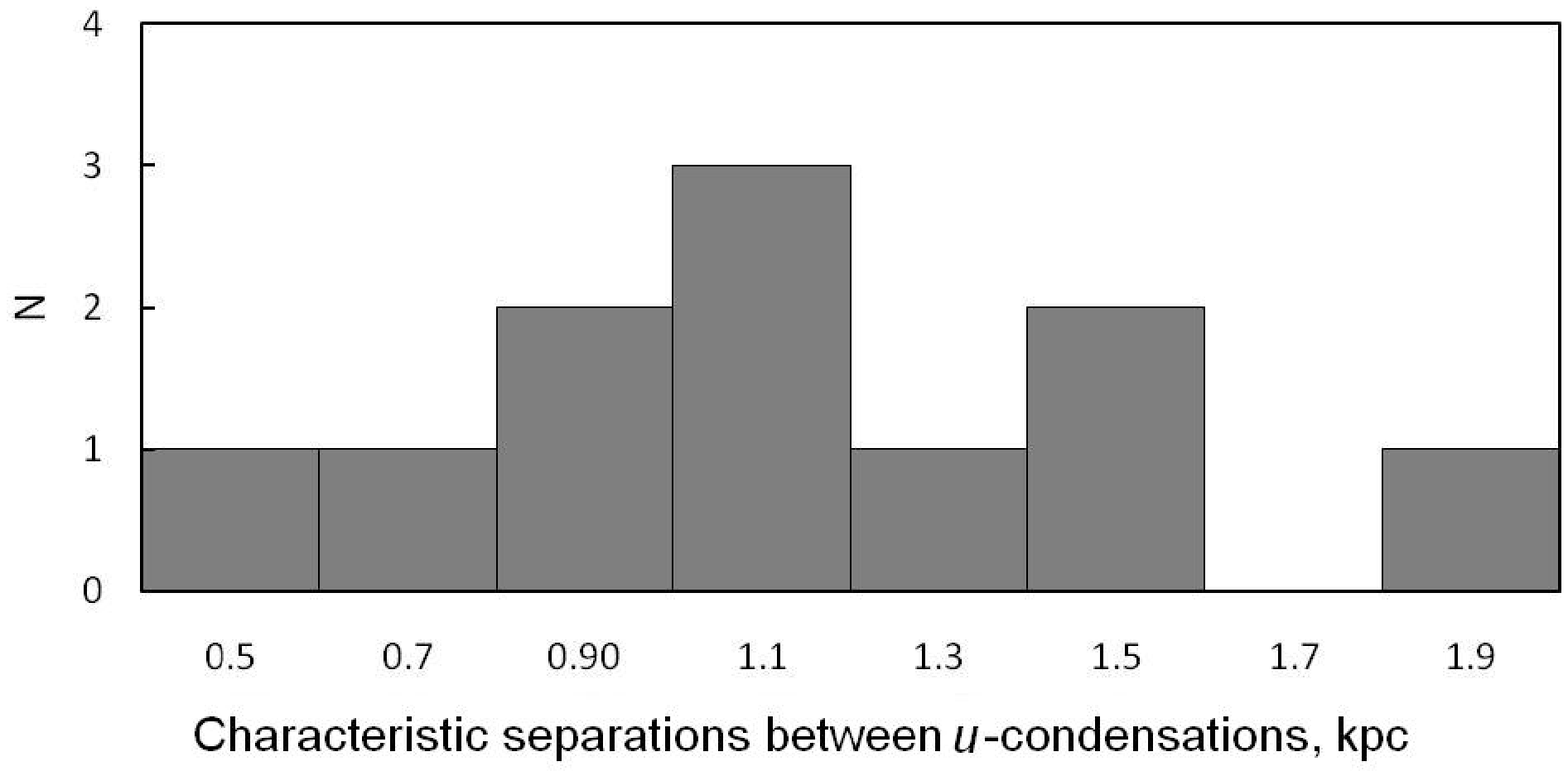}

   }

   \caption{The distributions of the separations between the adjacent clumps: the left -- H$\alpha$, the right -- $u$-band.}

\label{clumpsep}

\end{figure*}

To estimate the characteristic separation between the clumps in the H$\alpha$ emission 
line and between young star complexes in the $u$-band, we deprojected the MaNGaL image 
of the galaxy in the H$\alpha$ and the SDSS image in the $u$-band. For this purpose, we 
rotated the original images to put the line of nodes along the horizontal direction, 
and then stretched them vertically by adopting the inclination $i = 65^{\circ}$, in 
accordance with the kinematic inclination derived by analyzing the two-dimensional 
ionized gas velocity field measured with the Fabry-Perot interferometer. We applied the 
same orientation angles to deproject both images, because the ionized gas observed by 
us lies in the same plane as do the stars -- this is suggested by the consistent 
kinematics of the gas and stars revealed by us through our spectroscopic long-slit
observations (Proshina et al. 2019). Figure~6 shows the deprojected images of NGC~4324 
in the H$\alpha$ emission line and in the $u$-band and the separations between the 
centers of the adjacent clumps extracted from these images. In the deprojected H$\alpha$ image it can be clearly seen that the gaseous ring under study is intrinsically 
elliptical, which is typical for resonance rings (Buta 1995), with the resonance
rings being predominantly elongated perpendicular to the bar -- the so-called R1-type 
ring. An interesting manifestation of symmetry is that the localization of the 
starforming regions in the ring is pairwise, a half-turn of the galaxy apart (in 
Fig.~6, left, these pairs are connected by the straight-line segments). 
Is this a manifestation of the dynamical impact of the bar on the gas compression in 
the ring and the onset of star formation in the clumps? Figure~7 presents the 
distributions of clump separations. For the image that records the star formation just 
begun (the map in the H$\alpha$ emission line), the separations between clumps are 
grouped to one value, 0.65 -- 0.7~kpc; the second maximum of the histogram (near 
1.3~kpc) probably corresponds to the double characteristic separation, or to a 
''missed'' clump. On the $u$-band map corresponding to ''older'' starforming regions 
this regularity disappears.

\begin{figure*}[p]

\centering

\includegraphics[width=12cm]{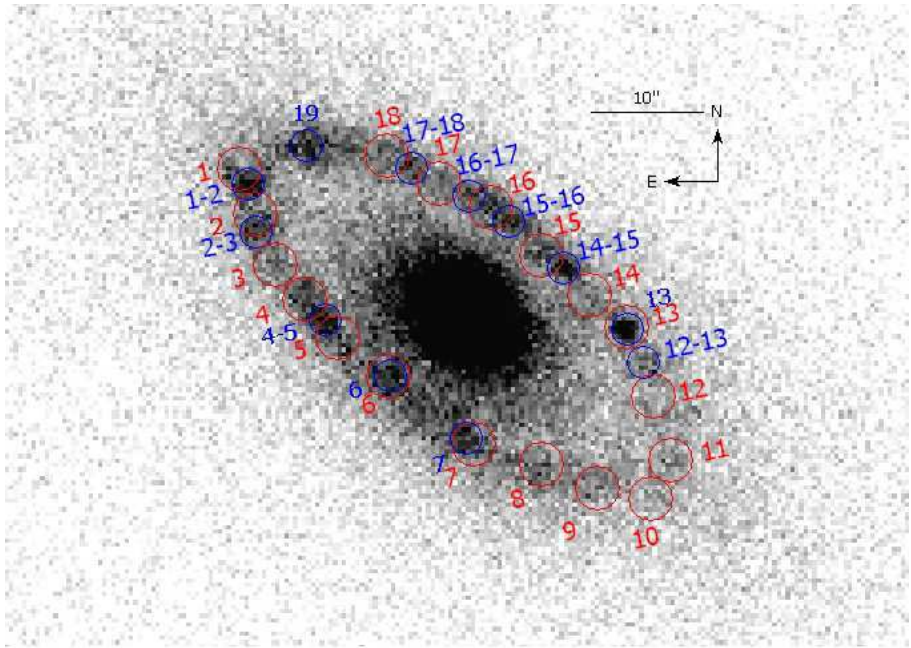}

\caption{The comparison of the star formation sites fixed in the emission line H$\alpha$ (red circles) and young star complexes fixed in the $u$-band (blue circles); the map in the H$\alpha$ is made through the MaNGaL data, the map in the $u$-band -- according to the SDSS/DR9 data.}

\label{u_ha}

\end{figure*}

\begin{figure*}[p]

\centering

   \vspace{0.5cm}

   \centerline{

   \includegraphics[width=0.45\textwidth]{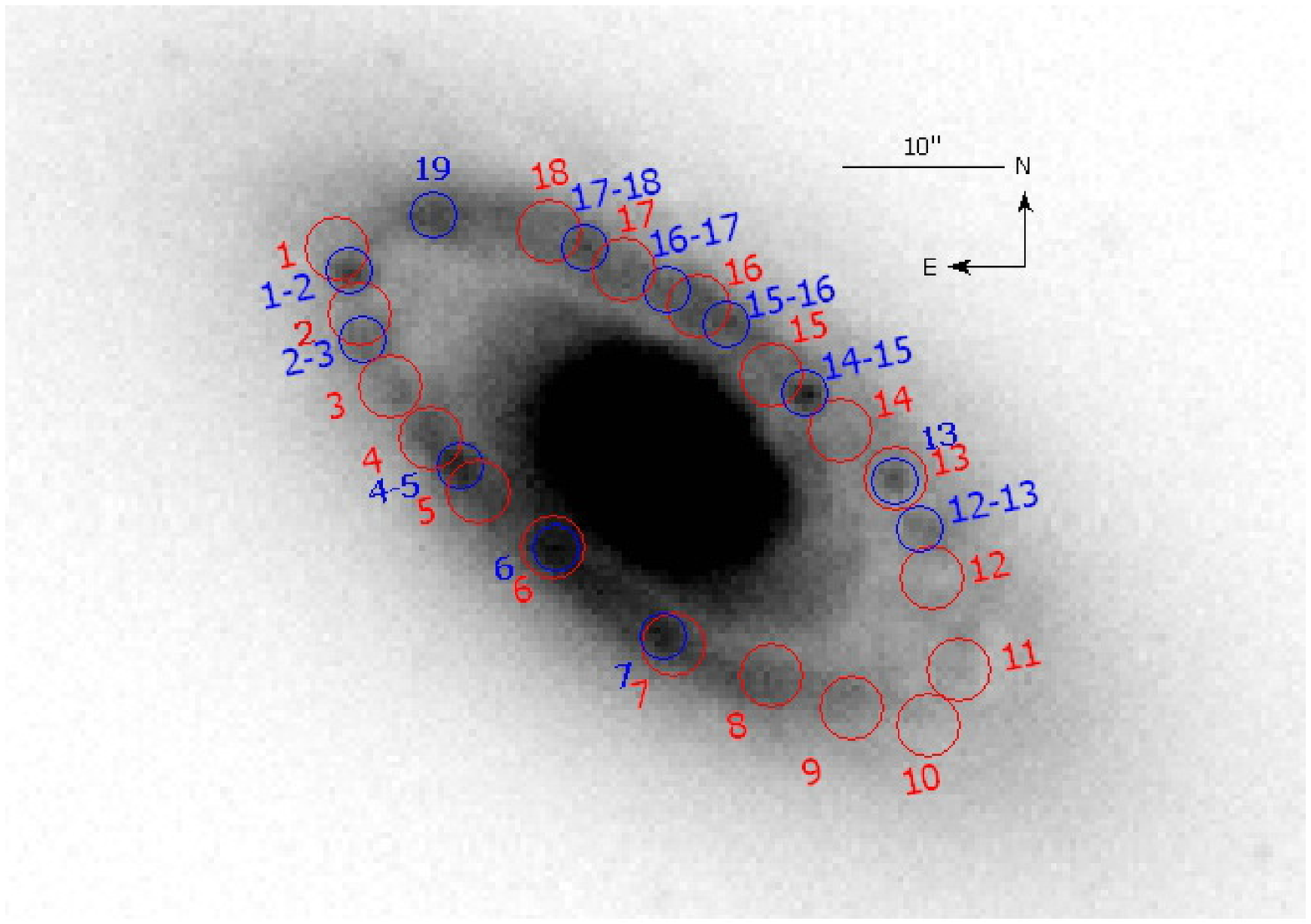}

   \includegraphics[width=0.45\textwidth]{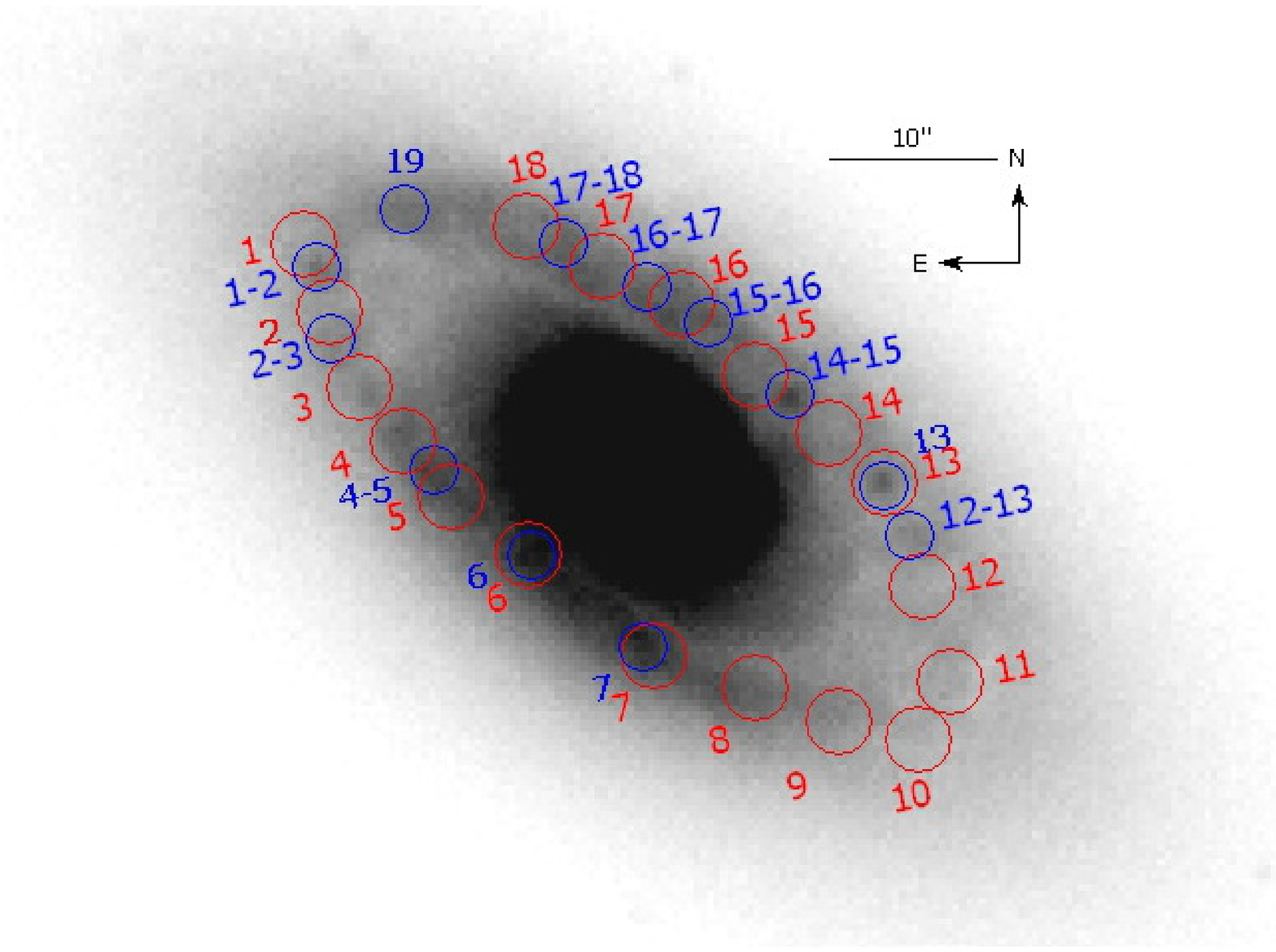}

   }

   \caption{The superposition of the star formation regions in the maps of NGC 4324 in the $g$-band (left) and in the $r$-band (right), fixed in the emission line H$\alpha$ (red circles), and young stellar complexes, fixed in the $u$-band (blue circles).}

\label{gr_uha}

\end{figure*}

Figures 8 and 9 show the SDSS images of the galaxy in the $u-$, $g-$, and $r-$ bands 
with the positions of the clumps found the $u$-band image (blue circles) and in the H$\alpha$ image (red circles) superposed on them. The clumps visible in the $u$-band are 
basically well-matured complexes of star clusters. It can be seen that only clumps 6, 
7, and 13 coincide on the H$\alpha$ and $u$-band maps. Many of the star complexes,
which are bright in the continuum, are visible in the gaps between H$\alpha$ clumps, 
for example, 1-2, 4-5, 12-13, 14-15, 15-16, 16-17, and 17-18; or at some offset from 
them, for example, complex 2-3; while complex 19, which has no analog in the narrow-band
H$\alpha$ and [NII]$\lambda$6583 images, is clearly seen in the images in both 
blue broad-band $u$ and $g$ filters. Recall that the most characteristic separation
between H$\alpha$ clumps found by us above is 0.7 kpc, while the characteristic 
separation between star complexes in the $u$-band is 1.05 and 1.5 kpc (Fig.~7). All of 
this taken together points to triggered star formation that occurs when the walls of 
the giant HI shells around young star complexes collide (Efremov and Elmegreen 1998; 
Egorov et al. 2015). Since the characteristic separation between clumps is different
for different wavelengths and since the ''visibility'' of starforming regions at 
different wavelengths is related to their age (the youngest ones are clearly seen in 
the H$\alpha$ emission line, the middle-aged ones are seen in the UV, and after 1--2 
Gyr we see the complexes to be bright in $u$ and $g$), there is the propagation of star 
formation along the ring that leads to a change in the characteristic separations 
between starforming complexes with time, as demonstrated by the histograms in Fig.~7. 
We cannot present an analogous histogram for the middle-aged starforming regions, 
because the GALEX space telescope observations in the NUV band had a spatial resolution 
of $6^{\prime \prime}$, which exceeds the sizes of the clumps and is comparable to the 
expected separations between them in the UV. Thus, in the GALEX data the structure of 
the starforming ring is washed out.

\section{DISCUSSION AND CONCLUSIONS}

Previously, a regular distribution of starforming regions has already been noted in the 
linear structures of disk galaxies. It is typical, for example, for tidal tails of 
interacting galaxies (Sotnikova and Reshetnikov 1998). It can be also encountered in 
spiral arms of galaxies (Efremov 2010; Gusev and Efremov 2013; Elmegreen et al. 2018). 
For instance, Efremov (2010) reported on star complexes having 0.6~kpc in size 
distributed like a chain along the north-western arm of the Andromeda galaxy with a 
typical spacing of 1.1~kpc, which, according to the author, reflects a regularity in 
the distribution of magnetic field lines. While investigating the regular chains of 
starforming complexes in the grand-design spiral galaxy NGC~628, Gusev and Efremov 
(2013) revealed characteristic separations between complexes that are a multiple 
of 0.4~kpc. Elmegreen et al. (2018) detected a regularity in the distribution of 
infrared clumps along filaments with a characteristic separation of 0.41 kpc in the 
dusty spiral galaxy M~100. In their recent paper, Gusev and Shimanovskaya (2020) noted 
such a regularity in the distribution of starforming regions along the resonance ring 
of the barred spiral galaxy NGC~6217 with a characteristic separation between
star complexes of 0.7~kpc. It is reported in the same paper that this is the first case 
when a regularity in the distribution of starforming regions is observed in ring 
structures. Now we see that this case is not unique. The ring in NGC~4324 may also be a 
resonance one: although no large-scale bar is seen in the galaxy, our isophotal 
analysis (Proshina et al. 2019), in particular, the jump in the ellipticity of 
isophotes at a radius of $13^{\prime \prime}- 15^{\prime \prime}$, points to a triaxial 
structure of the central part of the galaxy elongated approximately along the minor 
axis of the isophotes. The regularity in the distribution of starforming complexes in 
the ring of NGC~4324 suggests that the physical star formation mechanisms over local 
scales are the same in spiral and lenticular galaxies, while the differences arise when 
we start the analysis of large-scale structures: in spiral galaxies the major star 
formation occurs in spiral density waves, while in gas-rich lenticular galaxies it is 
organized into ring structures (Pogge and Eskridge 1993; Salim et al. 2012).

Since we definitely diagnose current and recent star formation in the ring of this 
lenticular galaxy and since the characteristic distance between the centers of clumps 
is 0.67~kpc, we can probably relate this size with the scale of inhomogeneities during 
the development of a gravitational instability  which leads also to star formation. It 
is of interest to estimate the critical gas surface density and to compare it with 
the observed one. It is found from theoretical calculations (Ledoux 1951) that the 
perturbations with the following wavelengths are unstable:

\begin{equation}
\label{toom}
\lambda _{crit}= \frac {2\pi ^2 G \Sigma _{gas}}{\kappa ^2}
\end{equation}
where $\kappa$ is the epicyclic frequency. From the ionized gas rotation curve 
calculated by us (Fig.~5), which is close to the circular velocity within the galactic 
potential due to the collisional nature of the gaseous subsystem, we see that the 
galaxy rotates rigidly up to the outer edge of the ring, $\sim 27^{\prime \prime}$. 
Under the assumption of rigid rotation, we obtain $\kappa \approx 1.7 \times 10^{-15}$ s$^{-1}$  in the ring. Taking $\lambda _{crit} = 0.67$~kpc, from Eq. (1) we then obtain $\Sigma ^{crit} _{gas} \approx 22\, M_{\odot}$~pc$^{-2}$. Let us now calculate the
observed molecular gas surface density in the ring by assuming, according to Alatalo et al. (2013), that the radial extent of the ring does not exceed $10^{\prime \prime}$ and $\log M(H_2)=7.97 \pm 0.02$ :

$$ \Sigma _{gas} ^{obs} = \frac{M_{H2}}{\pi(R^2-r^2)} \approx 4.3  \frac{M_{\odot}}{pc^2}. $$

Since the observed gas surface density is found to be lower than its critical value
and since we still observe star formation in the ring, this implies that, first, apart 
from the molecular gas, there is neutral hydrogen in the ring which we neglected in our 
calculations due to the lack of information about its quantitative content precisely in 
the ring, and, second, some additional factors leading to a gravitational instability 
and the onset of star formation are in action. One of the additional factors may be the 
so-called feedback from starforming regions, namely, gas compression by the giant 
gaseous shells expanding away from the starforming regions. It is also interesting that 
the parts of the ring where the ionized gas emission-line clumps show excitation by 
young stars are close in their azimuthal position to the ends of the oval structure 
(Fig.~10). Such a configuration may be associated with the so-called ansae, i.e.,
the regions of enhanced brightness at the ends of the bar. Although it is pointed out 
in the study by Martinez-Valpuesta et al. (2007) that current star formation proceeds 
very rarely in ansae, but for example, in NGC~4151, which, as NGC~4324, has no bar, but 
has an oval at the center, the ansae exhibit the blue colour and the H$\alpha$ emission line. 

We can estimate the mass of the star complexes using the SDSS images in the $u-$ and $g-$ bands as well as the mass-to-light calibrations from Bell et al. (2003). 
Since the ring is the inner one, the contribution of the underlying disk should be
taken into account in the aperture photometry of the starforming regions. For example, 
clump 13 is, in our view, the most suitable object for estimating the mass by the 
method described above, because this clump is the youngest and the most compact, as it 
was seen from the above analysis. We measured the fluxes in the $u-$ and $g-$ bands, 
converted them to magnitudes, and corrected them for the extinction in our Galaxy using 
the extinction coefficients from the NED for the SDSS photometric bands: $A_u = 0.102$ 
and $A_g = 0.08$. For the clump 13 the colour was found to be $(u- g) = 0.19$. Next, 
using Table~7 from Bell et al. (2003), we determine the ratio $M/L_g = 0.75$ and the 
mass of the young star complex of $7 \times 10^6\,M_{\odot}$. Several more largest
clumps (7, 1-2, 14-15) also have a stellar mass of $\sim 10^7\,M_{\odot}$. Such a mass 
of the star complexes is consistent with their size of 0.5~kpc while considering the gravitational instability of a gas (Cowie 1981).

\begin{figure*}[p]

\centering

\includegraphics[width=10cm]{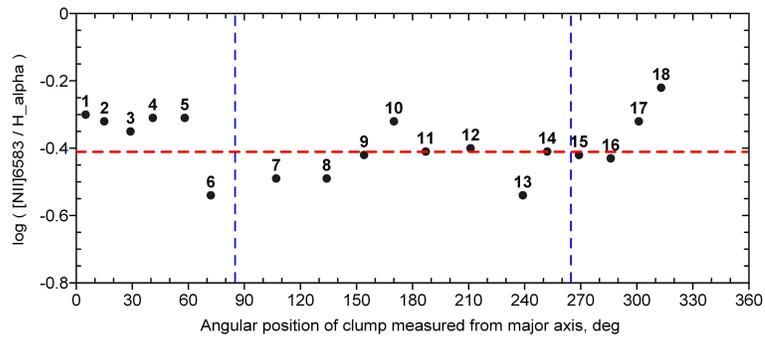}

\caption{Distribution of the ratio of the [NII]$\lambda$6583 and H$\alpha$ emission-line fluxes in azimuth along the ring. The typical error of the logarithm of the flux ratio is 0.01. The angle indicated on the horizontal axis is measured from the major axis of the isophotes, the north-eastern ending, counterclockwise. The horizontal dashed line demarcates the purely photoionization gas excitation (below the line) from the excitation with an admixture of shock waves. The vertical dashed lines indicate the probable orientation of the triaxial structure at the galactic center.}

\label{exci_azimuth}

\end{figure*}

The problem on the origin of the gas observed in the ring of this galaxy remains open: 
whether it is the gas that was returned by evolved stars of the galaxy or it has an 
accretional (external) origin. The consistent kinematics of the stellar and gaseous 
components argues for the first assumption. The mechanisms for the transfer of evolved 
gas from the galactic center to the periphery are considered in Marinacci et al. (2010). As regards the interaction with the environment, in the paper by Morales et al. (2018), which is devoted to searching for tidal features in nearby galaxies, NGC~4324 
was assigned to the galaxies in which no such features were detected. However, there is 
also another variant of accretion -- minor mergers.
If a gas-rich satellite of the galaxy had an orbital spin aligned with its rotation and 
if it fell in the plane of the stellar disk of NGC~4324, then this pattern of motion of 
the satellite could lead to gas accumulation in the disk of the galaxy under study 
without any apparent signatures of interaction.
The inflowing gas is accumulated in the ring having a resonant nature associated with 
the rotation of the triaxial structure at the galactic center (whose presence is 
suggested by the circumstantial evidence given above). Moreover our additional analysis 
of the gas line-of-sight velocity map in the H$\alpha$ line from Sil'chenko et al. (2019) presented above shows the presence of HII regions located far from the ring 
and rotating in the same plane and with velocities that more or less correspond to the 
galactic rotation at these radii. Our long-slit spectroscopic study (Proshina et al. 2019) showed clumps of emission in the H$\alpha$ and [N II] lines at great distances 
from the center, up to 12 kpc, with the gas velocities lying on a ''plateau'', i.e., 
coinciding with the velocity of the main galactic disk. At the same time, deviations 
from the circular flat-disk rotation model are also observed for three of the four 
outer HII regions that do not lie on the major axis of the isophotes (Fig. 5, middle). 
This most likely points to some inclination of the outer orbits of gaseous clouds -- to 
a warp of the gaseous disk probably associated with the capture of material from a 
plane that is not exactly coplanar with the stellar disk. Among the outer H$\alpha$ 
regions, region A (Fig.~5) is also seen in the continuum in the galaxy's blue images 
from SDSS with an absolute magnitude $M_g = -9.6$; it is probably an irregular gas-rich 
satellite of NGC~4324. According to our MaNGaL data, the nitrogen to H$\alpha$ line 
ratio allows the oxygen abundance in the gas to be estimated: with the calibrations 
from Pettini and Pagel (2004) and Marino et al. (2013) this estimate is $12+\log \mbox{(O/H)} = 8.44 \pm 0.04$, i.e., half the solar one and lower than the estimates for the outer NE and SW HII regions belonging to the galactic disk ($\sim 8.56$).

All these observational evidences confirm the hypothesis about the possible feeding of 
the disk in the lenticular galaxy with gas through the infall of gas-rich satellites 
and/or giant clouds. This aligned pattern of accretion of the satellites contributes to 
the star formation in the accreted gas, as was noted previously in Sil'chenko et al. (2019). Clumps are formed in the ring due to the gravitational instability, in which 
star formation ignits. The subsequent star formation triggers in the gaseous ring are 
probably the shock waves from evolving complexes of massive OB stars -- the first 
formed clusters of young stars in the gaseous clumps. In addition, the infall of a 
satellite or a giant gas cloud onto the galactic disk can serve as a trigger of another 
starburst. Thus, the chain of ''gaseous clumps -- star complexes'' observed by us is
a chain of the propagation of star formation both in space (in the ring) and in time. 
To clarify the question about the origin of the gas, we need a detailed mapping of 
NGC~4324 in the 21-cm HI line for both the galaxy itself and its environment.

\section{ACKNOWLEDGMENTS}

The study of the star-forming rings in S0 galaxies was supported by the Russian 
Foundation for Basic Research (project no. 18-02-00094). O.K. Sil'chenko also thanks 
the Interdisciplinary Scientific and Educational School of the Moscow State University 
''Fundamental and Applied Space Research''. In our work we used data from the NASA/IPAC 
Extragalactic Database (NED) operated by the Jet Propulsion Laboratory of the 
California Institute of Technology under contract with NASA and data the Lyon--Meudon 
HyperLEDA database. For our analysis we also invoked data from the GALEX and WISE space
telescopes. The NASA GALEX data were taken from the Mikulski Archive for Space 
Telescopes (MAST). The WISE data used by us were taken from the NASA/IPAC archive operated by the Jet Propulsion Laboratory of the California Institute of Technology under contract with the National Aeronautics and Space Administration.

\end{document}